\documentclass[ preprint, superscriptaddress,]{revtex4-2}
\usepackage[utf8]{inputenc}
\usepackage{CJKutf8}
\DeclareRobustCommand{\Chinese}[1]{%
  \begin{CJK*}{UTF8}{gbsn}#1\end{CJK*}%
}

\usepackage{graphicx}
\usepackage{dcolumn}
\usepackage{bm}
\usepackage{mathrsfs}
\usepackage{amsmath}
\usepackage[unicode=true,bookmarks=true]{hyperref}
\hypersetup{
    bookmarksopen=true,
    bookmarksnumbered=true,
    bookmarksdepth=3
}

\begin{document}

\preprint{APS/123-QED}

\title{Anomalous Hall enhancement in fully compensated non-collinear antiferromagnetic Cr-doped \texorpdfstring{$\mathrm{Mn}_{3}\mathrm{Sn}$}{Mn3Sn}}

\author{Ying Zhang} 
\thanks{These authors contributed equally to this work.}
\affiliation{%
 Department of Electrical and Computer Engineering, National University of Singapore, 117576 Singapore 
}%
\affiliation{%
 National University of Singapore (Chongqing) Research Institute, Chongqing Liang Jiang New Area, Chongqing 401123, China
}%

\author{Xin Chen}
\thanks{These authors contributed equally to this work.}
\altaffiliation{%
Present address: School of Electrical and Electronic Engineering, Nanyang Technological University, Singapore 639798, Singapore
}%
\affiliation{%
 Department of Electrical and Computer Engineering, National University of Singapore, 117576 Singapore
}%

\author{Hang Xie}
\affiliation{%
 State Key Laboratory of Spintronics, Nanjing University, Suzhou 215163, China
}%
\affiliation{%
 School of Integrated Circuits, Nanjing University, Suzhou 215163, China
}%

\author{Ziyan Luo}
\affiliation{%
 School of electronic information, Central South University, Changsha, 410004, China
}%

\author{Yihong Wu}
\email{elewuyh@nus.edu.sg}
\affiliation{%
 Department of Electrical and Computer Engineering, National University of Singapore, 117576 Singapore
}%
\affiliation{%
 National University of Singapore (Chongqing) Research Institute, Chongqing Liang Jiang New Area, Chongqing 401123, China
}%

\author{Lei Shen \Chinese{(沈雷)}}
\email{shenlei@nus.edu.sg}
\affiliation{%
 Department of Mechanical Engineering, National University of Singapore, 117575 Singapore
}%
\affiliation{%
 National University of Singapore (Chongqing) Research Institute, Chongqing Liang Jiang New Area, Chongqing 401123, China
}%

\date{\today}

\begin{abstract}
The non-collinear antiferromagnet (AFM), $\mathrm{Mn}_{3}\mathrm{Sn}$, exhibits ultrafast spin dynamics and negligible stray fields, making it a promising candidate for fast and energy-efficient spintronic devices. However, its practical applications remain limited by the relatively small Hall response compared with ferromagnets. Here, we demonstrate through both first-principles calculations and experiments that Cr-doping is a promising strategy to enhance the anomalous Hall transport in $\mathrm{Mn}_{3}\mathrm{Sn}$ without breaking the antiferromagnetic order. 
Our first-principles calculations reveal an effective Cr-doping window that enhances the anomalous Hall conductivity while largely preserving the host coplanar inverse triangular AFM order. 
Berry curvature and Wannier Hamiltonian analyses further reveal that Cr substitution amplifies the pre-existing Berry curvature hotspots through coordinated changes in the energy alignment, local interband gap, and Mn-derived orbital weight of the relevant bands.
Experimentally, we have found that Cr doping not only supports our computational predictions (enhancement of the anomalous Hall conductivity), but also suppresses the transition to helical or spin-glass phases at low temperatures. 
These findings demonstrate that Cr-doping is an effective strategy for stabilizing the AFM order while simultaneously improving the anomalous Hall transport properties over a wide temperature range in $\mathrm{Mn}_{3}\mathrm{Sn}$.
\end{abstract}

\maketitle


\section{\label{sec:level1}INTRODUCTION}
Future spintronic devices demand magnetic platforms that are compatible with high density integration, fast operation, and reduced power consumption~\cite{tsaiPicosecondUltralowpowerSwitching2026,zhouFieldfreeFullSwitching2026a}. Non-collinear antiferromagnets (AFMs) have emerged as a key platform in this area due to their unique advantages over ferromagnets (FMs), including negligible stray fields and ultrafast dynamics~\cite{rimmlerNoncollinearAntiferromagneticSpintronics2025,baltzAntiferromagneticSpintronics2018,daldinAntiferromagneticSpintronics2024a}. In contrast to conventional collinear antiferromagnets, their non-collinear spin textures can break the relevant symmetries for the anomalous Hall effect (AHE)~\cite{guoLargeAnomalousNernst2017,chenAnomalousHallEffect2014a}, anomalous Nernst effect (ANE)~\cite{panGiantAnomalousNernst2022,reichlovaImagingWritingMagnetic2019,ikhlasLargeAnomalousNernst2017}
even in the absence of net magnetic moments. Among these, the AHE provides a direct transport readout of antiferromagnetic order for device application without requiring complex heterostructure design~\cite{qinRoomtemperatureMagnetoresistanceAllantiferromagnetic2023}. 

Originally established in ferromagnets~\cite{nagaosaAnomalousHallEffect2010a}, the intrinsic AHE is now also observed in certain non-collinear AFMs where both time-reversal symmetry $\mathcal{T}$ and its combination with fractional translation $\mathcal{T}_{\tau}$ are broken, enabling a net anomalous Hall conductivity (AHC). In contrast, collinear AFMs typically exhibit negligible AHE because the $\mathcal{T}_{\tau}$ symmetry remains preserved. The non-collinear antiferromagnetic $\mathrm{Mn}_{3}\mathrm{Sn}$ is a prototypical example, in which the inverse triangular (IT) magnetic order generates a sizable room temperature AHC, making it an ideal platform for exploiting the interplay between magnetic symmetry, electronic structure and electrical transport~\cite{gorenAntiferromagneticLocalVolatile2025,chenAnomalousTransportDue2021}. However, the practical implementation of $\mathrm{Mn}_{3}\mathrm{Sn}$ remains limited by the restricted stability range of the IT phase and the strong sensitivity of its Hall response to magnetic phase, stoichiometry, and film growth conditions~\cite{sungMagneticPhaseDependence2018,xieMagnetizationSwitchingPolycrystalline2022b,taylorAnomalousTopologicalHall2020,gasMagneticPhaseDiagram2025}. The IT order in $\mathrm{Mn}_{3}\mathrm{Sn}$ is stable only within a relatively narrow temperature window, $275$--$420~\mathrm{K}$ in bulk (often even narrower in thin films). Below $\sim 275~\mathrm{K}$, the IT phase transitions into helical or spin-glass states, causing the AHC to diminish significantly. Therefore, a central challenge is to further enhance the AHC while maintaining a robust Hall response over a broader temperature rangeand preserving the compensated noncollinear antiferromagnetic framework.

Several strategies have therefore been explored to tune the Hall response and magnetic phase stability of $\mathrm{Mn}_{3}\mathrm{Sn}$, among which compositional engineering through chemical doping provides a particularly direct route because it addresses two central objectives simultaneously. First, doping can shift the Fermi level, $E_{\mathrm{F}}$, relative to the Weyl-node-derived Berry-curvature hot spots located slightly below $E_{\mathrm{F}}$ in stoichiometric $\mathrm{Mn_{3}Sn}$~\cite{kublerWeylFermionsAntiferromagnetic2018}. Second, it can suppress or eliminate the low-temperature spin-reorientation transition ($\sim260~\mathrm{K}$), which drives the IT state toward a helical/spin-glass phase and strongly suppresses the AHE below $\sim50~\mathrm{K}$. 
The earliest and most extensively investigated example is Mn-excess self-doping, $\mathrm{Mn}_{3+x}\mathrm{Sn}_{1-x}$. Even a modest $\sim1\%$ variation in Mn concentration shifts the temperature of the ANE/AHE maximum by nearly $50~\mathrm{K}$, consistent with a shift of $E_{\mathrm{F}}$ relative to the Weyl points~\cite{chenAnomalousTransportDue2021}. At larger off-stoichiometry, epitaxial films exhibit emergent Kondo-gap behavior and resonance-enhanced THz responses~\cite{khadkaKondoPhysicsAntiferromagnetic2020}, while thin films with bulk off-stoichiometric compositions display both sign reversal and enhancement of the spin Hall conductivity~\cite{bangarLargeSpinHall2023}. These observations highlight the strong sensitivity of transport properties to carrier concentration and band filling. 
Substitution on the Sn site has also been explored such as Ga, Si, and Ge, with varying effects on the Hall response and magnetic phase stability~\cite{wenEnhancedAnomalousHall2025,rajuTunableHighNeel2025a,yanoTuningAnomalousHall2024b}. 
In contrast, direct substitution on the magnetic Mn sublattice can produce more pronounced changes in both the Hall and magnetic responses, although often at the cost of substantial magnetic reconstruction. 
Fe doping provides a representative example: magnetic bubbles and biskyrmions emerge in $\mathrm{Mn}_{2.1}\mathrm{Fe}_{0.9}\mathrm{Sn}$~\cite{liuSpontaneousMagneticBubbles2020}; coexistence of ferromagnetic and antiferromagnetic phases with strong exchange bias appears in $\mathrm{Mn_{2}FeSn}$~\cite{sapkotaSynthesisPropertiesStoichiometric2025a}; and Fe-induced uniaxial anisotropy stabilizes nontrivial low-temperature spin textures in $\mathrm{Mn}_{2.75}\mathrm{Fe}_{0.25}\mathrm{Sn}$~\cite{lowTuningTopologicalProperties2022}. More recently, combined theoretical and experimental studies further associated Fe-doped $\mathrm{Mn_{3}Sn}$ with a higher-order-exchange-driven canted noncoplanar state and an additional scalar-spin-chirality contribution to the Hall response~\cite{singhHigherOrderExchange2024}. Together, these studies suggest that although Fe substitution can effectively tune transport properties, it frequently introduces competing magnetic complexity. 
Against this backdrop, Cr substitution has emerged as a promising route for improving the Hall response of $\mathrm{Mn}_{3}\mathrm{Sn}$ while relatively preserving its antiferromagnetic framework. Chen \textit{et al.} experimentally demonstrated that Cr doping suppresses the spin-reorientation transition and enhances the AHC in polycrystalline $\mathrm{Mn_{3}Sn}$ films, with modulation-doped samples reaching an AHC of $184.8~\Omega^{-1}\mathrm{cm}^{-1}$ at $50~\mathrm{K}$, in sharp contrast to the nearly vanishing response of undoped films at the same temperature~\cite{chenGiantEnhancementAnomalous2021}. However, this study mainly focused on experimental observation, and the microscopic origin of the enhancement remained unresolved. Yano \textit{et al.} further showed through systematic single crystal studies that Cr provides more reproducible, although somewhat weaker, low-temperature AHE enhancement than Fe, and that stable AHE down to 2K can ultimately be achieved through Fe/Cr co-doping in $\mathrm{Mn}_{3}\mathrm{Sn}$ ~\cite{yanoTuningAnomalousHall2024b}. While providing important insights into magnetic and transport evolution, their work did not directly address the intrinsic electronic mechanism responsible for Cr-induced AHC enhancement. Despite these advances, the microscopic mechanism by which Cr substitution enhances the AHC in $\mathrm{Mn}_{3}\mathrm{Sn}$ remains unclear, particularly regarding its impact on the underlying electronic structure and Berry curvature distribution.

Here, we present a combined first-principles and experimental investigation of Cr-doped $\mathrm{Mn}_{3}\mathrm{Sn}$ that provides a microscopic picture of AHC enhancement while largely preserving the host non-collinear AFM framework. We first perform first-principles calculations with virtual crystal approximation (VCA) to identify an optimal low-concentration Cr doping window for achieving large AHC in $\mathrm{Mn}_{3}\mathrm{Sn}$. By combining Berry curvature and Wannier Hamiltonian analyses, we resolve how Cr substitution systematically modifies the key electronic factors governing the intrinsic AHC, leading to an amplification of the pre-existing Berry curvature hotspots. Using explicit supercell calculations, we further reveal how discrete local Cr configurations modify the AHC response while largely preserving the host in-plane AFM order. To test these predictions, we synthesize Cr-doped $\mathrm{Mn}_{3}\mathrm{Sn}$ thin films and characterize their magnetic and transport properties. The measurements demonstrate enhanced AHC near room temperature and improved low-temperature Hall retention, qualitatively consistent with the theoretically identified enhancement trend.


\section{RESULTS AND DISCUSSION}
\subsection{
\texorpdfstring
{Pristine \ensuremath{\mathrm{Mn}_{3}\mathrm{Sn}}}
{Pristine Mn3Sn}
}
\label{sec:level2}
\begin{figure}[!tp]
\centering 
\includegraphics[width=\columnwidth]{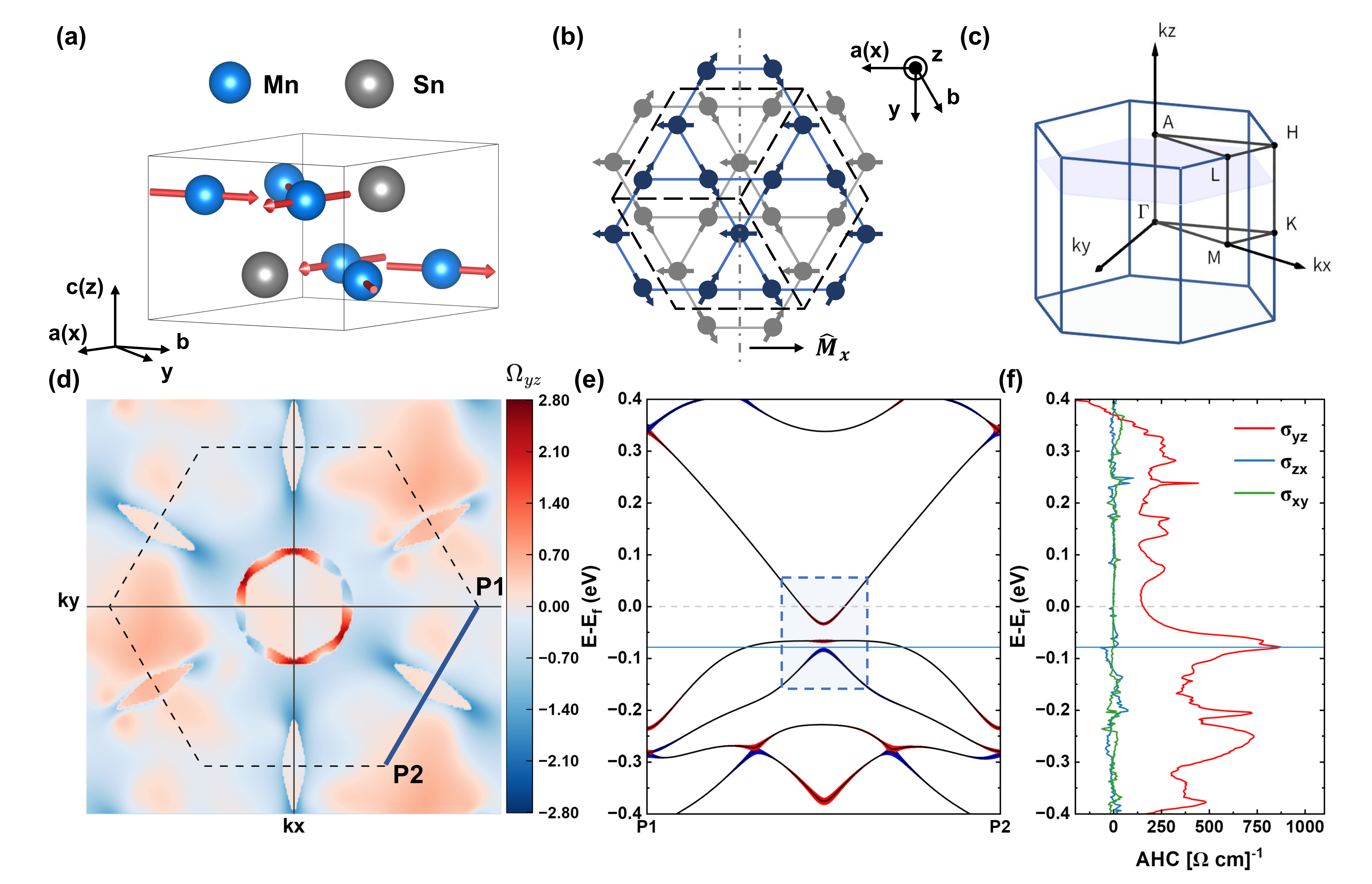} 
\caption{(a) Crystal structure of $\mathrm{Mn}_{3}\mathrm{Sn}$. (b) Magnetic structure of the Mn spins in type-I $\mathrm{Mn}_{3}\mathrm{Sn}$ viewed in projection down the $c$ axis. The gray dashed lines indicate the $\hat{M}_{x}$ mirror plane of the symmetry operation $\{\hat{M}_{x}\mid\boldsymbol{\tau}\}$. (c) First Brillouin zone of $\mathrm{Mn}_{3}\mathrm{Sn}$. (d) Berry curvature map on the $k_{z}=0.27$ plane at Fermi energy. The color scale represents $\operatorname{sgn}(\Omega_{yz})\log_{10}(|\Omega_{yz}|+1)$, with red and blue denoting positive and negative BC, respectively. (e) Local Berry curvature band across the hotspot region P1--P2 marked in (d), with the line width scaled by $\log_{10}(|\Omega_{yz}|+1)$. (f) Energy dependent anomalous Hall conductivity of type-I $\mathrm{Mn}_{3}\mathrm{Sn}$ along three different directions.}
\label{fig:fig1}
\end{figure}

We first establish pristine $\mathrm{Mn}_{3}\mathrm{Sn}$ as the host system whose anomalous response is already encoded in its magnetic symmetry and electronic structure. $\mathrm{Mn}_{3}\mathrm{Sn}$ crystallizes in a hexagonal layered structure with AB stacking in the $\mathrm{P}6_{3}/\mathrm{mmc}$ space group (No. 194), where Mn and Sn occupy the $6h$ and $2c$ Wyckoff positions, respectively [Fig.~\ref{fig:fig1}(a)]. For the $\boldsymbol{k}=0$ non-collinear antiferromagnetic order with in-plane $120^{\circ}$ spin arrangement, symmetry analysis yields four distinct magnetic configurations~\cite{horiAntiferromagneticFerromagneticTransition1990,sohGroundstateMagneticStructure2020}, summarized in Supporting Information SI1~\cite{SM}. Among them, only configurations I and III support finite intrinsic AHC, with symmetry-allowed tensor components $\sigma_{yz}$ and $\sigma_{zx}$, respectively. This tensor selectivity follows directly from magnetic symmetry. For example, configuration I [Fig.~\ref{fig:fig1}(b)] preserves the nonsymmorphic symmetry $\{\hat{M}_{x}\mid\boldsymbol{\tau}\}$, where $\hat{M}_{x}$ is the mirror operation with respect to the $yz$ plane and $\boldsymbol{\tau}=(0,0,c/2)$ is a half lattice translation along the $c$ axis. Because the intrinsic AHC is given by the Brillouin zone integral of Berry curvature [Eq.~(\ref{eq:ahc})], the transformation of the Berry curvature (BC) pseudovector under this symmetry enforces $\sigma_{xy}=\sigma_{zx}=0$, leaving $\sigma_{yz}$ as the only finite component, in agreement with our first-principles calculations. Configuration III is related to configuration I by an in-plane $90^{\circ}$ spin rotation and correspondingly supports only $\sigma_{zx}$. Total energy calculations identify type I as the lowest energy configuration among the four candidates [Table~S1], and unless stated otherwise the following discussion focuses on type I [Fig.~\ref{fig:fig1}(b)].

To investigate its microscopic origin, we first examined the $\Omega_{yz}$ projected band structure along the conventional high-symmetry path [Fig.~S1(b)], where $\Omega_{yz}$ denotes the Berry curvature component associated with $\sigma_{yz}$ [Eq.~(\ref{eq:bc})]. 
Although several enhanced contributions are visible near the Fermi energy ($E_{\mathrm{F}}$), the dominant hotspot responsible for the AHC maximum is not fully resolved on the high symmetry path, indicating a more general distribution of BC hotspots in the three-dimensional Brillouin zone. A systematic search of the full Brillouin zone identifies the strongest hotspot on the $k_{z}=0.27$ plane (in units of reciprocal lattice vectors; illustrated in Fig.~\ref{fig:fig1}(c)) at $E_{\mathrm{F}}$, and the corresponding BC map is shown in Fig.~\ref{fig:fig1}(d). Rather than passing directly through the core of the underlying three-dimensional BC hotspot, this section intercepts only its adjacent outer flank. Consistently, the local band structure along the in-plane P1--P2 path crossing this region [Fig.~\ref{fig:fig1}(e)] reveals a very narrow avoided crossing located close to but not exactly at $E_{\mathrm{F}}$, with opposite-sign BC accumulated on the two sides of the small gap. Such narrowly avoided crossings near $E_{\mathrm{F}}$ are widely recognized as one of the characteristic microscopic sources of strong BC driven Hall responses~\cite{nagaosaAnomalousHallEffect2010a,xiaoBerryPhaseEffects2010a}. This naturally suggests a low-disturbance compositional tuning strategy that introduces Cr to optimize the existing BC hotspots while preserving the host non-collinear magnetic order.

\subsection{\label{sec:level3}Origin of the Hall enhancement by uniform Cr doping}
\begin{figure}[!htbp] 
\centering 
\includegraphics[
    width=\textwidth,
    height=0.65\textheight,
    keepaspectratio
]{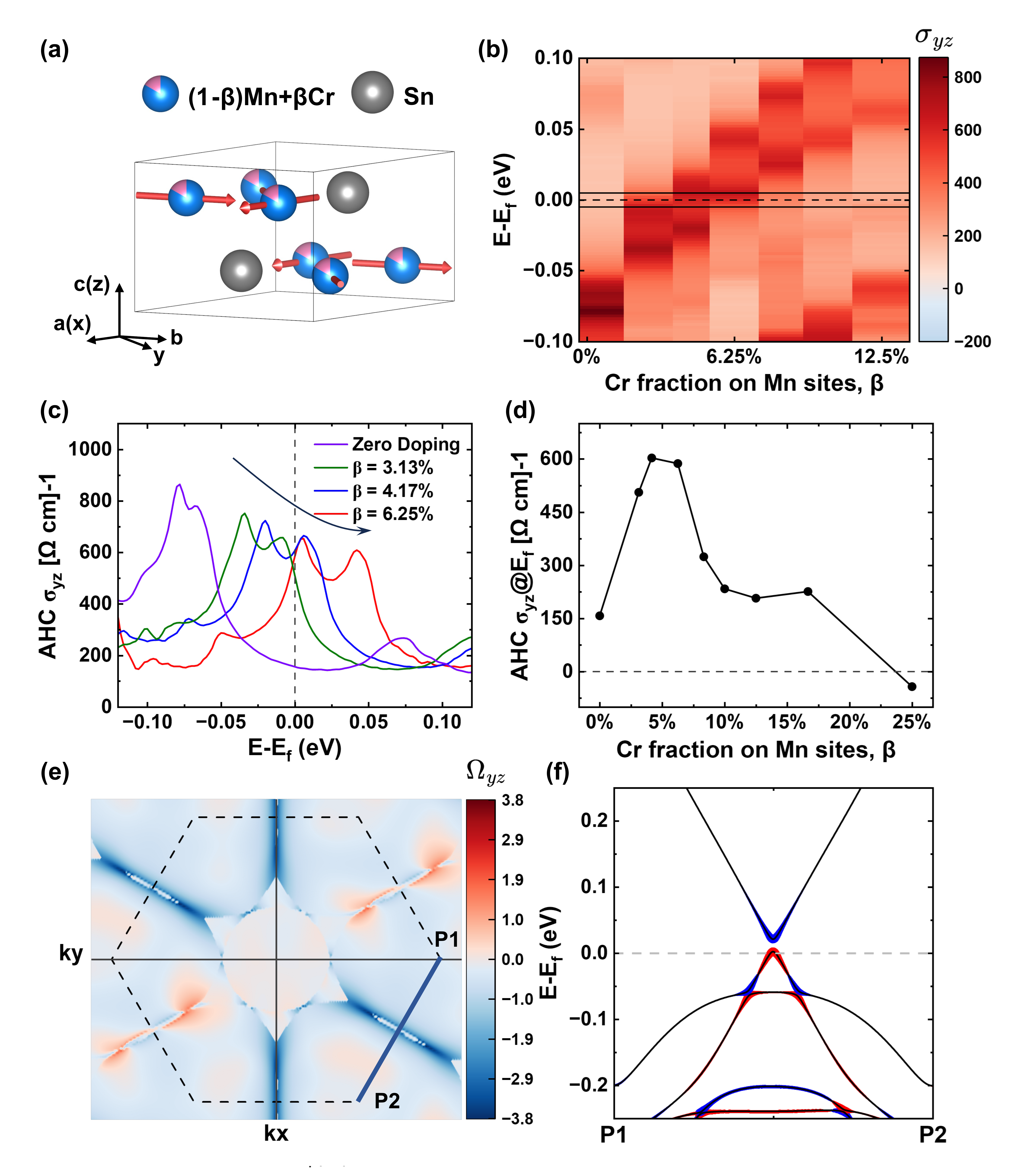} 
\caption{\label{fig:fig2}Virtual crystal approximation (VCA) results for the anomalous Hall conductivity (AHC) and Berry curvature (BC) hotspot in Cr-doped $\mathrm{Mn}_{3}\mathrm{Sn}$. (a) Schematic of the VCA model for Cr substitution on the Mn sublattice. (b) Energy and doping ratio dependent AHC $\sigma_{yz}$ pseudocolor plot of Cr doped $\mathrm{Mn}_{3}\mathrm{Sn}$. The dashed line represents the neutral filling (Fermi level), while the solid black line is employed to ascertain the error bar, which differs by $5~\mathrm{meV}$ from the Fermi level. (c) Energy dependent AHC curves within optimal concentration window. (d) AHC $\sigma_{yz}$ at the Fermi level as a function of Cr concentration. (e) BC map for $\beta=4.17\%$ at the Fermi level on the $k_{z}=0.355$ plane, with the color scale given by $\operatorname{sgn}(\Omega_{yz})\log_{10}(|\Omega_{yz}|+1)$. (f) Local BC band across the hotspot region P1--P2 marked in (e), with the line width scaled by $\log_{10}(|\Omega_{yz}|+1)$.} 
\end{figure}

Within the Kubo Berry curvature framework [Eq.~(\ref{eq:ahc})], the intrinsic AHC is given by the Brillouin zone integral of the BC over occupied states, so its magnitude depends sensitively on how BC hotspots aligned with the Fermi level ($E_{\mathrm{F}}$). From the BC expression [Eq.~(\ref{eq:bc})], large contributions are expected when band pairs near $E_{\mathrm{F}}$ are separated by small gaps and coupled by sizable velocity matrix elements. To isolate the average alloy effect of Cr substitution, we first employ the virtual crystal approximation, in which Cr substitution on the Mn sublattice is represented by a spatially averaged alloy potential [Fig.~\ref{fig:fig2}(a)]. Within this description, low-concentration Cr substitution opens a clear enhancement window for the AHC while largely preserving the host coplanar non-collinear magnetic framework. The VCA models are described by the nominal composition $\mathrm{Mn}_{3-x}\mathrm{Cr}_{x}\mathrm{Sn}$, where the Mn site Cr fraction is defined as $\beta=x/3$ [Eq.~(\ref{eq:vca_potential})]. For example, $\beta=4.17\%$ corresponds to $x=0.125$. To capture both the experimentally relevant low-concentration regime and the onset of stronger perturbation to the host magnetic state, $\beta$ was varied from 0 to 25\%. Computational details are provided in the Methods section.

We first examine the stability of the magnetic structure under Cr doping. As summarized in Sec. SI2, the Fermi energy shifts systematically with increasing Cr concentration, whereas the in-plane RMSE, which quantifies the deviation of the spin orientations from the pristine IT configuration, local magnetic moments, and total cell moment vary only modestly over the VCA range considered, indicating that Cr doping acts primarily as a weak perturbation while preserving the in-plane IT order in the pristine $\mathrm{Mn}_{3}\mathrm{Sn}$. To further test the robustness of the coplanar magnetic state against out of plane perturbation, the self-consistent canting stability calculations are carried out for $\mathrm{Mn}_{2.75}\mathrm{Cr}_{0.25}\mathrm{Sn}$ ($\beta=8.33\%$) and $\mathrm{Mn}_{2.5}\mathrm{Cr}_{0.5}\mathrm{Sn}$ ($\beta=16.7\%$) (see Supplemental Information SI3~\cite{SM}). Starting from initial out-of-plane canting angles ranging from $0^{\circ}$ to $30^{\circ}$, all calculations converged back to states very close to the coplanar limit, and the minimum total energy was consistently obtained at zero canting. These results indicate that, within the present VCA description, low and intermediate level Cr substitution does not stabilize an out-of-plane magnetic component, in contrast to the noncoplanar canted tendencies reported for Fe substituted or electron doped $\mathrm{Mn}_{3}\mathrm{Sn}$~\cite{singhHigherOrderExchange2024}.

Having the magnetic robustness of the host IT order, we analyze the electronic and Hall response. Because the magnetic symmetry remains unchanged in the average alloy picture, $\sigma_{yz}$ remains the only nonzero AHC tensor component, as shown in Fig.~\ref{fig:fig1}(f). Around the optimal Cr fraction of $\beta \approx 5\%$, the overall band dispersion is largely preserved, and the dominant change is a local energy shift of the band near the Fermi level rather than a reconstruction of the global electronic structure (see Fig.~S7 in SI4). Correspondingly, the AHC peak under Fermi energy found in pure $\mathrm{Mn}_{3}\mathrm{Sn}$ gradually moves toward the Fermi level while increasing Cr concentration [Fig.~\ref{fig:fig2}(b, c)]. As $\beta$ approaches and exceeds $8.3\%$, the overall band modification becomes stronger, the peak broadens and disperses, with the dominant Hall channel no longer further enhanced. The evolution of AHC at $E_{\mathrm{F}}$ directly reflects this trend. In Fig.~\ref{fig:fig2}(d), an optimal concentration window appears around $\beta=4.17\%$, where the AHC at $E_{\mathrm{F}}$ reaches $603~\mathrm{S\,cm^{-1}}$, corresponding to a about $400\%$ enhancement over pristine $\mathrm{Mn}_{3}\mathrm{Sn}$. Further increasing the doping concentration to $\beta \approx 10\%$, the AHC decreases to approximately $200~\mathrm{S\,cm^{-1}}$ due to the reconstruction of the electronic structure and eventually reverses sign at $\beta=25\%$.

The same full $k$ space BC analysis was then performed for the optimally doped VCA system with $\beta=4.17\%$ to determine the microscopic origin of the Hall enhancement. The results show that $4.17\%$ Cr concentration does not generate a new Hall channel or qualitatively reconstruct the BC distribution, but instead mainly realigns the existing hotspot structure of the $\mathrm{Mn}_{3}\mathrm{Sn}$ $\sigma_{yz}$ channel toward the Fermi level. Fig.~\ref{fig:fig2}(e, f) illustrates a representative hotspot plane selected from the strongest BC response of the doped system at $E_{\mathrm{F}}$. At the optimal Cr fraction of $\beta=4.17\%$, the BC distribution on the $k_{z}=0.355$ plane closely resembles that of pristine $\mathrm{Mn}_{3}\mathrm{Sn}$ on the $k_{z}=0.344$ plane at its AHC peak energy ($E_{\mathrm{max}}$) [Fig.~S8(a)], indicating that $4.17\%$ Cr doping preserves the $\mathrm{Mn}_{3}\mathrm{Sn}$ BC hotspots while bringing it into much closer alignment with the Fermi level. A direct comparison of the corresponding local band structures [Fig.~\ref{fig:fig2}(f) and Fig.~S8(d)] further shows that the overall band shape remains unchanged, while only the relative energy position of the near-crossing feature shifts. In this way, the optimal $\beta=4.17\%$ Cr substitution reduces the energy difference between the $E_{\mathrm{F}}$ and the $E_{\mathrm{max}}$ from $0.078~\mathrm{eV}$ to $0.02~\mathrm{eV}$. Additional comparisons on other corresponding representative planes of pristine and doped $\mathrm{Mn}_{3}\mathrm{Sn}$ are provided in Supporting Information SI6~\cite{SM}.

\begin{figure}[!htbp] 
\centering 
\includegraphics[
    width=\textwidth,
    height=0.65\textheight,
    keepaspectratio
]{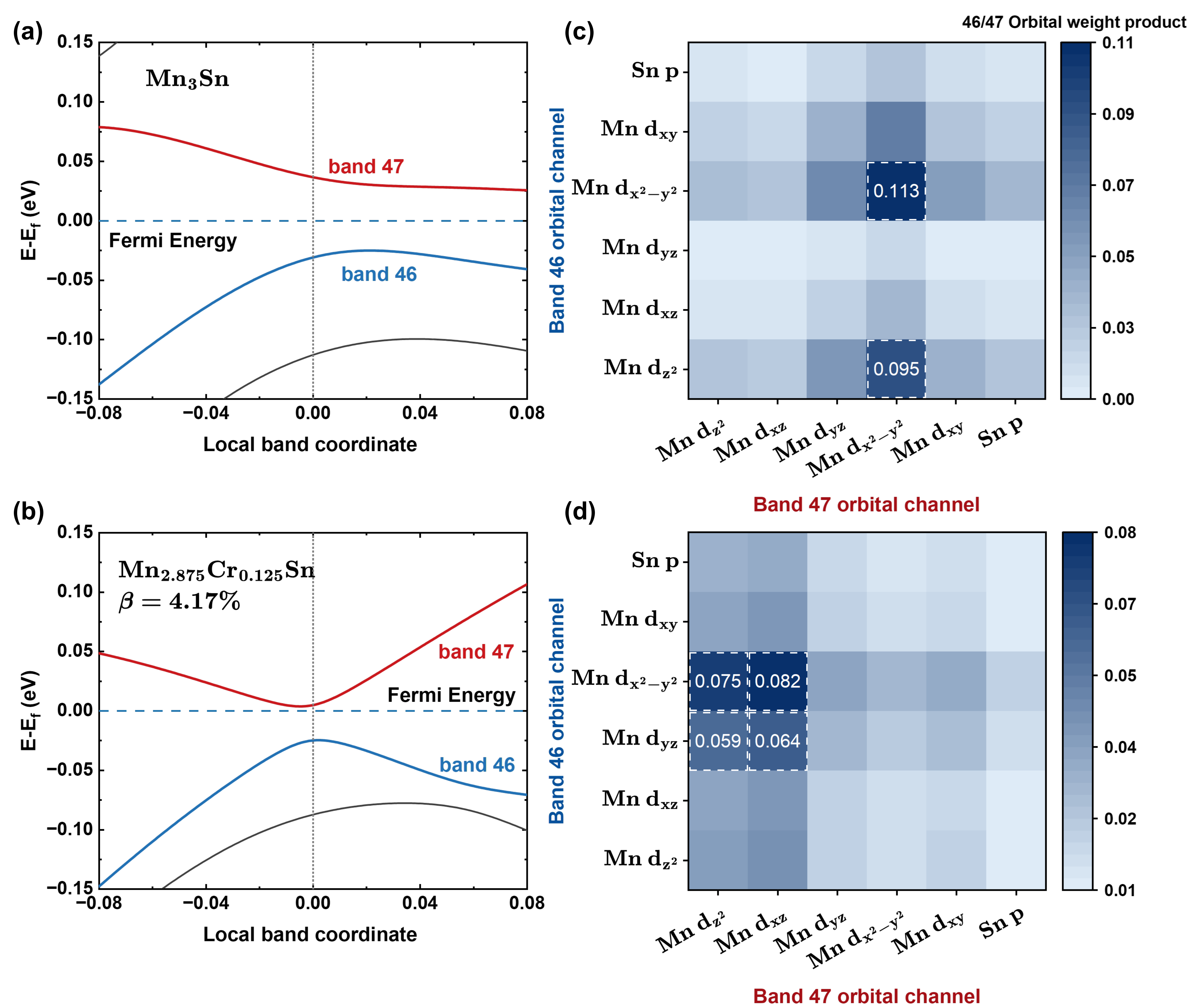} 
\caption{\label{fig:fig3}Wannier Hamiltonian origin of anomalous Hall enhancement in $\mathrm{Mn}_{2.875}\mathrm{Cr}_{0.125}\mathrm{Sn}$. Representative local AHC hotspot analysis for pristine $\mathrm{Mn}_{3}\mathrm{Sn}$ (a, b) and $\mathrm{Mn}_{2.875}\mathrm{Cr}_{0.125}\mathrm{Sn}$ with $\beta=4.17\%$ (c, d), identified by Kubo motivated screening of Wannier interpolated small gap band pairs near $E_{\mathrm{F}}$. (a, c) The local band structure around the selected hotspots, centered at $\boldsymbol{k}_{0}=(0.333,0.000,0.250)$ for $\mathrm{Mn}_{3}\mathrm{Sn}$ and $\boldsymbol{k}_{0}=(0.333,0.000,0.375)$ for  $\beta=4.17\%$, respectively, with the doped case exhibiting a reduced local gap at $E_{\mathrm{F}}$. (b, d) The corresponding Wannier channel maps for the 46/47 band pair, obtained from the product of the projected Wannier weights of the two bands, revealing a redistribution of the dominant orbital mixing contributions upon Cr doping. For doping result (d), the ``Mn'' in labels denotes the VCA Mn/Cr mixed site.} 
\end{figure}

The Wannier Hamiltonian $H_{ij}(\mathbf{R})$ was decomposed into local and intersite contributions to resolve the underlying orbital-level mechanism. In the Wannier construction, the basis was restricted to the Mn like $d$ manifold and the Sn $p$ orbitals. The local terms primarily exhibit smooth background renormalization upon Cr substitution. A gradual increase in the $d$-orbital onsite energies is observed on the virtual Mn/Cr sites, while the corresponding same-site $d$ orbital spin mixing strength is reduced, with much weaker average changes in the Sn $p$ blocks [Fig.~S11]. This behavior suggests that the AHC enhancement is not driven by a sharp anomaly in the local terms and instead points to the intersite hopping network that governs the BC hotspots near $E_{\mathrm{F}}$. We therefore analyzed matrix elements connecting Wannier orbitals on different atomic sites, including both $\mathbf{R}=0$ and $\mathbf{R}\neq 0$, through the momentum space Hamiltonian 
\begin{equation} 
H_{ij}(\boldsymbol{k})
=
\sum_{\mathbf{R}}
e^{i\boldsymbol{k}\cdot\mathbf{R}}
H_{ij}(\mathbf{R}).
\label{Ham}
\end{equation} 

A Kubo-inspired local velocity proxy, 
\begin{equation}
v_{\mathrm{proxy}}^{nm}(\boldsymbol{k})
=
\left|
\frac{
2\,\operatorname{Im}
\left[
v_{x}^{nm}(\boldsymbol{k})
v_{y}^{mn}(\boldsymbol{k})
\right]
}{
\left[
E_{m}(\boldsymbol{k})-E_{n}(\boldsymbol{k})
\right]^{2}
+\eta^{2}
}
\right|,
\label{velocity}
\end{equation} 
was defined to screen the BC like hotspots associated with occupied–unoccupied band pairs separated by small local gaps where $v_{\alpha}^{nm}(\boldsymbol{k})$ are the velocity matrix elements obtained from the Wannier interpolated Hamiltonian, $E_{m}(\boldsymbol{k})-E_{n}(\boldsymbol{k})$ is the local band gap and $\eta$ is a finite broadening parameter introduced to regularize the energy denominator. For $\mathrm{Mn}_{3}\mathrm{Sn}$, the dominant 46/47 hotspot appears near $\boldsymbol{k}_{0}=(0.333,0.000,0.250)$, with a gap of $0.067~\mathrm{eV}$ and $v_{\mathrm{proxy}}=45.9$[Fig.~\ref{fig:fig3}(a)]. While for $\beta=4.17\%$, the corresponding hotspot shifts to $\boldsymbol{k}_{0}=(0.333,0.000,0.375)$, accompanied by a gap reduction to $0.027~\mathrm{eV}$ and a strong increase of $v_{\mathrm{proxy}}$ to $1012.5$, nearly 22 times the pristine value[Fig.~\ref{fig:fig3}(b)]. Simultaneously, the dominant Wannier channel changes from Mn-derived $d_{x^{2}-y^{2}}$--$d_{x^{2}-y^{2}}$ intersite mixing to Mn-derived $d_{x^{2}-y^{2}}$--$d_{z^{2}}$ intersite mixing [Fig.~\ref{fig:fig3}(c, d)]. Further increasing $\beta$ to $8.33\%$, the same 46/47 channel remains identifiable but its $v_{\mathrm{proxy}}$ is largely reduced, consistent with the weakening of the AHC beyond the optimal concentration. Additional analysis is summarized in Supplementary Information SI7~\cite{SM}.

Within the VCA framework, Berry curvature and Wannier Hamiltonian analyses show that dilute Cr substitution amplifies the pre-existing Berry curvature hotspots of $\mathrm{Mn}_{3}\mathrm{Sn}$ through a coordinated modification of the relevant electronic structure. The Berry curvature comparison reveals a favorable energy shift of the original hotspot, while the local Hamiltonian analysis further identifies gap narrowing and redistribution of the Mn-derived orbital weight in the associated band pair. Together, these changes enhance the corresponding interband Kubo contribution and account for the increased intrinsic AHC. However, the VCA provides only an idealized, composition-averaged description and does not explicitly capture local chemical environments, configuration-dependent symmetry breaking, or impurity-scattering effects. To complement this averaged picture at the atomic level, we next examine representative Cr configurations using explicit supercell calculations and evaluate their local magnetic structures, electronic states, and AHC tensor responses.

\subsection{\label{sec:level4}Local Cr doping effect and symmetry dependent Hall response}
\begin{figure}[!htbp] 
\centering 
\includegraphics[
    width=\textwidth,
    height=0.65\textheight,
    keepaspectratio
]{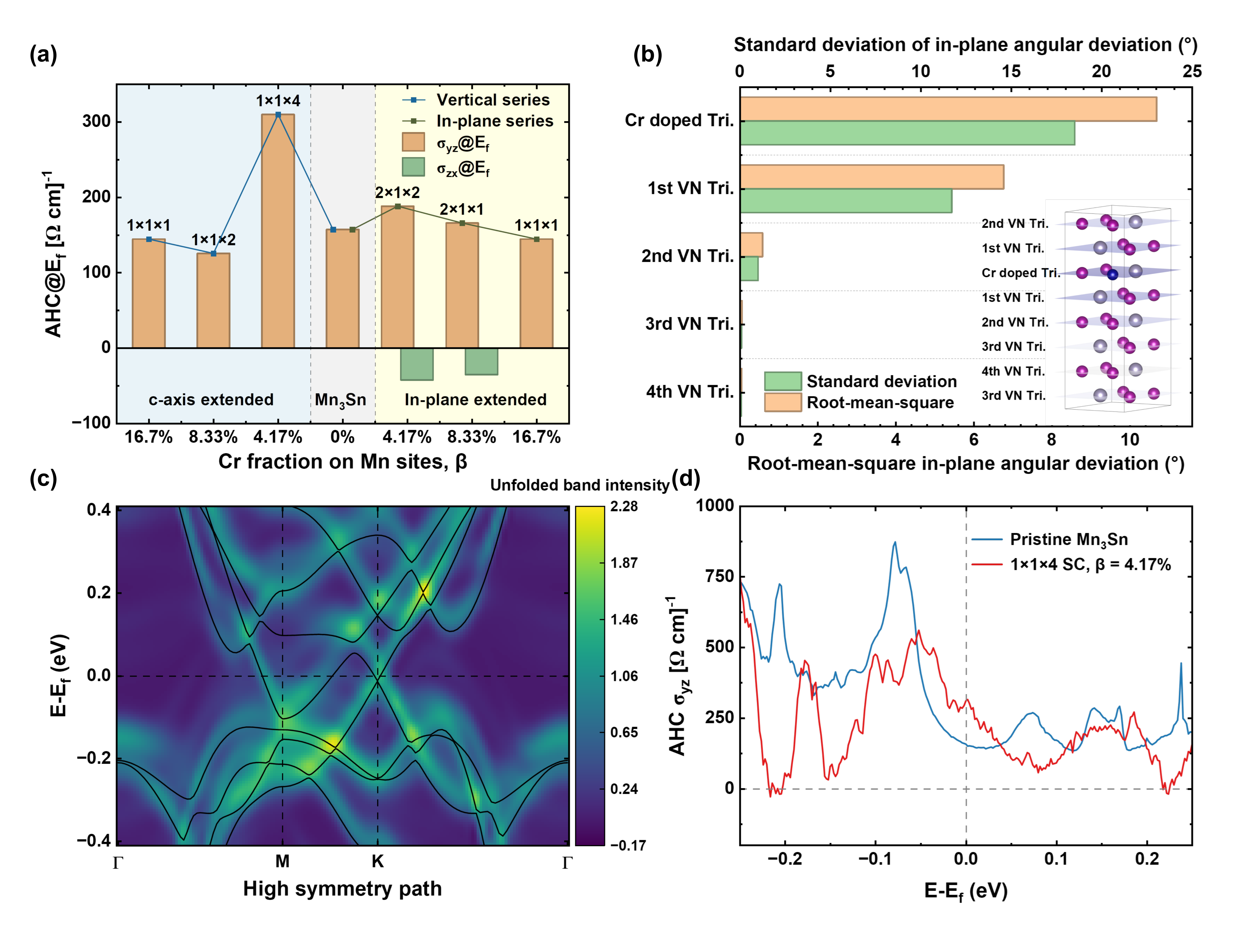} 
\caption{\label{fig:fig4}The Supercell configuration on the anomalous Hall response of Cr doped $\mathrm{Mn}_{3}\mathrm{Sn}$. (a) Calculated anomalous Hall conductivity at the Fermi level for representative supercell configurations constructed by two different expansion schemes. The blue and yellow shaded regions denote the $c$-axis extended and in-plane extended series, respectively. (b) Summary of the magnetic perturbation in the $1\times1\times4$ supercell (SC) at $\beta=4.17\%$. Inset is the schematic structure of $1\times1\times4$ supercell with one Mn replaced by Cr. The VN denotes the vertical neighbor Mn triangles relative to the Cr containing triangle, while 1st and 2nd indicate the corresponding neighbor order. (c) Unfolded band structure intensity of the $1\times1\times4$ supercell. The band solid line shows the reference band structure of host $\mathrm{Mn}_{3}\mathrm{Sn}$. (d) Energy dependent anomalous Hall conductivity $\sigma_{yz}$ for the $1\times1\times4$ supercell compared with $\mathrm{Mn}_{3}\mathrm{Sn}$.} 
\end{figure}

Representative $c$-axis and in-plane-expanded supercells were constructed, where the notation $a \times b \times c$ denotes the replication factors along the three lattice vectors, with one Mn atom replaced by Cr in each supercell.
Figure~\ref{fig:fig4}(a) compares several representative supercells with different Cr distributions. 
Because the VCA describes Cr substitution through a composition-averaged potential whereas the supercell approach treats specific atomic configurations, quantitative agreement in the absolute AHC values is not necessarily expected. The more meaningful comparison is therefore the concentration-dependent tendency. Despite their smaller AHC values, both the $c$-axis and in-plane-expanded series exhibit a similar dilute Cr enhancement trend, indicating that the enhancement identified by VCA persists across the representative explicit configurations. 
The $c$-axis extended series most closely follows the VCA trend, as the Hall response remains dominated by the parent $\sigma_{yz}$ component, and the $1\times1\times4$ supercell at $\beta=4.17\%$ [see inset of Fig.~\ref{fig:fig4}(b)] shows the strongest enhancement near $E_{\mathrm{F}}$, reaching nearly twice the value of pristine $\mathrm{Mn}_{3}\mathrm{Sn}$. The corresponding magnetic perturbation is also highly localized. For the $1\times1\times4$ supercell, the in-plane angular deviations of the Mn triangles decay rapidly away from the Cr-containing layer and become negligible beyond the second neighboring layer [Fig.~\ref{fig:fig4}(b)]. No sizable out-of-plane moment is generated, with the maximum canting angle remaining below $0.28^{\circ}$. The unfolded band structure likewise retains the overall $\mathrm{Mn}_{3}\mathrm{Sn}$ dispersion, with only localized deviations near Cr-derived orbital contributions [Fig.~\ref{fig:fig4}(c) and Fig.~S13(c)]. The energy dependent AHC curve further shows that explicit Cr substitution shifts the dominant AHC peak toward $E_{\mathrm{F}}$, producing the enhanced $\sigma_{yz}$ response [Fig.~\ref{fig:fig4}(d)]. The original AHC peak near $-0.078~\mathrm{eV}$ becomes lower and broader, reflecting a redistribution of the BC contribution over a wider energy range due to the explicit supercell configuration rather than the emergence of a distinct Hall mechanism. Supercells involving in-plane expansion exhibit a different tensor character due to the reduced local symmetry associated with the in-plane Cr distribution. The strongest $\sigma_{yz}$ response within this group is still found at $\beta=4.17\%$, although the enhancement is weaker and accompanied by a finite $\sigma_{zx}$ component. This indicates that the low-concentration optimum remains identifiable, while dopant arrangement redistributes the Hall response across tensor channels. Nevertheless, the magnetic perturbation remains limited, with the largest out-of-plane canting angle still below $0.34^{\circ}$, and the host coplanar AFM framework is still largely preserved (Fig.~S14 in SI9). Structural relaxations of representative supercells confirm that the Cr-induced perturbation remains short localized, with enhanced in-plane distortions but no stable out-of-plane magnetic component (Sec. SI10). 

Although the present supercells do not constitute a full configurational average, the representative $c$-axis and in-plane-expanded configurations consistently show that the Cr-induced magnetic perturbation remains highly localized and does not generate an appreciable out-of-plane magnetic component. The persistence of the dilute Cr AHC enhancement across these distinct local arrangements further suggests that these representative configurations capture the main configurational trends relevant to the present analysis, while the variation in AHC magnitude and tensor components reflects the local configurational sensitivity omitted by the VCA. However, these supercell calculations still describe the intrinsic AHC and do not include impurity scattering or other disorder related transport effects. We therefore next examine whether the overall tendency toward AHC enhancement at relatively low Cr incorporation is also observed experimentally in sputtered thin films under more realistic conditions.

\subsection{\label{sec:level5}Experimental validation}

Two types of Cr-doped $\mathrm{Mn}_{3}\mathrm{Sn}$ films, namely uniform and modulation-doped samples, were prepared by magnetron sputtering, and their structural, magnetic, and electrical properties were systematically characterized by X-ray diffraction (XRD), magnetometry, and Hall transport measurements over the temperature range of $50$--$300~\mathrm{K}$.
For the modulation-doped samples, we deposit $\left[\mathrm{Mn}_{3}\mathrm{Sn}(t_{1})/\mathrm{Cr}(t_{2})\right]_{N}$ multilayers and use the individual layer thicknesses to control the equivalent Cr composition; here $t_{1}$ and $t_{2}$ are the thicknesses of the $\mathrm{Mn}_{3}\mathrm{Sn}$ and Cr sublayers in $\mathrm{nm}$, and $N$ is the number of repetitions. 
The undoped $\mathrm{Mn}_{3}\mathrm{Sn}$ film was prepared with the same total thickness of $72~\mathrm{nm}$, corresponding to a $2.4 \times 30~\mathrm{nm}$ stacking configuration, and served as the control reference.
Uniform doping, denoted as MS50CrX, is achieved by co-sputtering $\mathrm{Mn}_{3}\mathrm{Sn}$ and Cr targets, with sputtering powers of $50~\mathrm{W}$ and $X~\mathrm{W}$ for the $\mathrm{Mn}_{3}\mathrm{Sn}$ and Cr targets, respectively. Based on calibrated deposition rates, the Cr incorporation of MS50Cr10 was estimated to be comparable to that of $\left[\mathrm{Mn}_{3}\mathrm{Sn}(2.4)/\mathrm{Cr}(0.15)\right]_{30}$. All samples were deposited on $\mathrm{SiO_{2}}/\mathrm{Si}$ substrates by room-temperature magnetron sputtering, followed by post-annealing to promote crystallization and dopant interdiffusion. Detailed deposition and characterization procedures are provided in the Methods section. Unless otherwise specified, the discussion below focuses on four representative samples: undoped $\mathrm{Mn_{3}Sn}$, $\left[\mathrm{Mn_{3}Sn}(2.4)/\mathrm{Cr}(0.15)\right]_{30}$, $\left[\mathrm{Mn_{3}Sn}(2.4)/\mathrm{Cr}(0.6)\right]_{30}$, and the co-doped sample MS50Cr10. These samples are selected to provide a systematic comparison between the undoped reference, different Cr incorporation levels, and different doping strategies.

Structural characterization confirms that both the uniformly doped and modulation doped films largely retain the $\mathrm{Mn}_{3}\mathrm{Sn}$ polycrystalline structure (Sec. SI11). The XRD patterns (Fig.~S15) remain broadly similar to that of undoped $\mathrm{Mn}_{3}\mathrm{Sn}$, with peak positions and extracted lattice constants close to the bulk values, indicating good Cr solubility and no evidence of major secondary phases or significant structural reconstruction. Small angle X-ray reflectivity measurements further show that the modulation doped series exhibits only weak or no clear superlattice periodicity except for the thickest Cr layer sample, suggesting full intermixing rather than $\mathrm{Mn}_{3}\mathrm{Sn}$/Cr multilayer as in the as-deposited state. 
Magnetic measurements further show that the representative samples remain in a comparable magnetization range at $300~\mathrm{K}$~\cite{zhaoMagneticPropertiesAnomalous2021,wangIntegrationNoncollinearAntiferromagnetic2019,markouNoncollinearAntiferromagneticMn3Sn2018} and that Cr incorporation does not introduce an anomalously large net moment (Fig.~S16 in SI12). More importantly, the temperature-dependent magnetization shows that low-concentration Cr doping suppresses the low-temperature spin-reorientation transition~\cite{sungMagneticPhaseDependence2018,taylorAnomalousTopologicalHall2020}, although this suppression weakens at higher Cr concentrations, where the system instead tends toward a different magnetic phase. 

\begin{figure}[!htbp] 
\centering 
\includegraphics[
    width=\textwidth,
    height=0.7\textheight,
    keepaspectratio
]{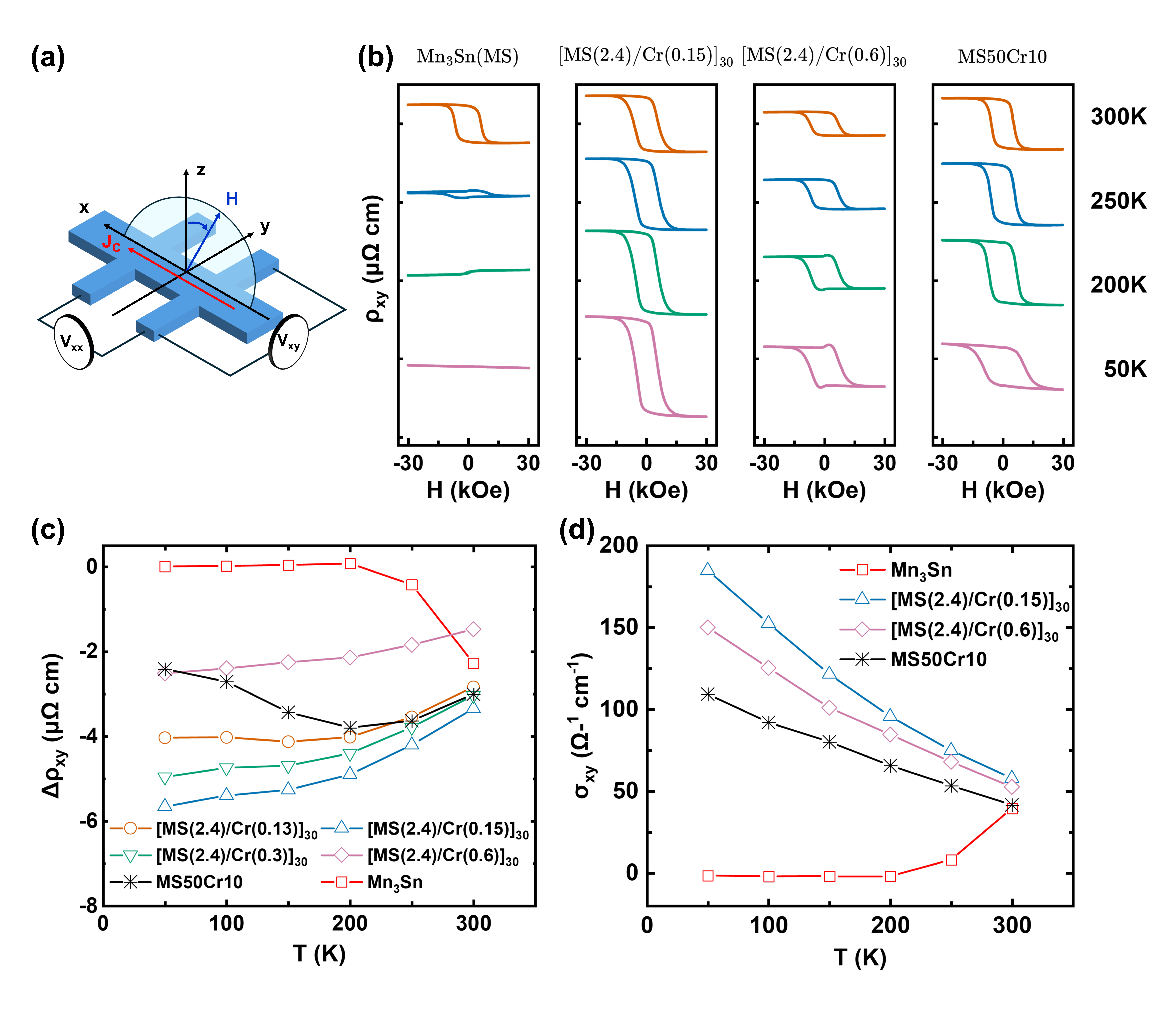} 
\caption{\label{fig:Fig5}(a) Schematic of Hall bar and electrical transport measurement. (b) Magnetic field dependence of anomalous Hall resistivity for $\mathrm{Mn}_{3}\mathrm{Sn}$, $\left[\mathrm{Mn}_{3}\mathrm{Sn}(2.4)/\mathrm{Cr}(t_2)\right]_{30}$ and MS50Cr10 at different temperatures. (c) Temperature dependence of the zero-field Hall resistivity $\Delta\rho_{xy}$ for selected representative samples. (d) Temperature dependence of zero field AHC for $\mathrm{Mn}_{3}\mathrm{Sn}$, $\left[\mathrm{Mn}_{3}\mathrm{Sn}(2.4)/\mathrm{Cr}(t_2)\right]_{30}$ and MS50Cr10.} 
\end{figure}

Figure~\ref{fig:Fig5} shows the electrical measurement configuration [Fig.~\ref{fig:Fig5}(a)] and corresponding transport measurement results. Figure~\ref{fig:Fig5}(b) presents representative Hall loops for the aforementioned four samples, while Fig.~\ref{fig:Fig5}(c) summarizes the temperature dependence of the corresponding zero-field Hall signal $\Delta\rho_{xy}$ for the selected representative samples and the modulation-doped series [Eq.~(6)], including MS50Cr10 as the uniformly doped reference; the full uniformly doped series is provided in Fig.~S17.
In the undoped sample (MS), a negative zero field anomalous Hall resistivity of about $-2.27~\mu\Omega\,\mathrm{cm}$ is observed at $300~\mathrm{K}$, consistent with reported polycrystalline $\mathrm{Mn}_{3}\mathrm{Sn}$ films~\cite{zhaoMagneticPropertiesAnomalous2021,ikedaAnomalousHallEffect2018,higoAnomalousHallEffect2018} and characteristic of the room-temperature non-collinear AFM state. Its rapid suppression upon cooling and near disappearance below $200~\mathrm{K}$ are likewise consistent with the known low temperature magnetic phase evolution of $\mathrm{Mn}_{3}\mathrm{Sn}$~\cite{sungMagneticPhaseDependence2018,taylorAnomalousTopologicalHall2020}. With the incorporation of Cr, a common trend becomes clear, with both modulation doped and uniformly doped samples exhibiting an optimum low Cr window in their Hall response, where the room temperature signal is enhanced and the low temperature signal is more effectively retained. In the modulation doped series with fixed $t_{1}=2.4~\mathrm{nm}$, samples with $t_{2}$ from $0.13$ to $0.3~\mathrm{nm}$ all display substantially larger $\Delta \rho_{xy}$ than MS, with $\left[\mathrm{Mn}_{3}\mathrm{Sn}(2.4)/\mathrm{Cr}(0.15)\right]_{30}$ giving the strongest overall response. Quantitatively, the zero field Hall resistivity of $\left[\mathrm{Mn}_{3}\mathrm{Sn}(2.4)/\mathrm{Cr}(0.15)\right]_{30}$ increases from $3.35~\mu\Omega\,\mathrm{cm}$ at $300~\mathrm{K}$ to $5.56~\mu\Omega\,\mathrm{cm}$ at $50~\mathrm{K}$. Further increasing $t_{2}$ to $0.6~\mathrm{nm}$ still yields a finite Hall signal that strengthens upon cooling, but its reduced magnitude and the increased low temperature magnetization in Fig.~S16(e) suggest a stronger perturbation of the host magnetic framework at higher Cr incorporation. 
A similar trend is observed in the fixed $t_{2}=0.15~\mathrm{nm}$ series [Fig.~S17], where reducing $t_{1}$ from $2.4$ to $1~\mathrm{nm}$ effectively increases the relative Cr incorporation and progressively weakens the Hall signal, eventually leading to near suppression. 

The uniformly doped low Cr sample MS50Cr10, also shows an enhanced Hall response at room temperature and a retained signal at intermediate to low temperatures, indicating that Cr induced Hall tuning is not exclusive to the modulation doped geometry. 
Its $300~\mathrm{K}$ $M$--$H$ hysteresis loops and $M$--$T$ measurements are comparable to those of the low Cr modulation doped sample, suggesting that both low Cr incorporation routes preserve the weak moment magnetic character of $\mathrm{Mn}_{3}\mathrm{Sn}$ without producing a large FM response. 
The main difference appears at the lowest temperatures, where the Hall signal of MS50Cr10 decreases rather than continuing to increase. This contrast indicates that the Hall response is not governed solely by the average Cr concentration, but can be further modulated by the spatial distribution and local environment of Cr atoms. This difference is also qualitatively consistent with the supercell calculations, which show that local Cr configuration modulates both the magnetic perturbation and the AHC. 
Since the actual Cr distribution after annealing is likely more complex than any single ordered supercell model, the comparison is best interpreted as evidence that local Cr configuration can modulate the Hall response beyond the average doping level.

The corresponding anomalous Hall conductivity, plotted in Fig.~\ref{fig:Fig5}(d), further clarifies the low doping Hall enhancement behavior observed experimentally. 
Here, the zero field anomalous Hall conductivity was evaluated as $\sigma_{xy}\approx-\rho_{xy}/\rho_{xx}^{2}$, where $\rho_{xx}$ is the longitudinal resistivity measured at $H=0$ (Sec. SI13). 
At $300~\mathrm{K}$, $\sigma_{xy}$ is already enhanced in the Cr doped samples relative to undoped MS, reaching $57.8$, $52.6$, and $41.66~\Omega^{-1}\,\mathrm{cm}^{-1}$ in $\left[\mathrm{Mn}_{3}\mathrm{Sn}(2.4)/\mathrm{Cr}(0.15)\right]_{30}$, $\left[\mathrm{Mn}_{3}\mathrm{Sn}(2.4)/\mathrm{Cr}(0.6)\right]_{30}$, and MS50Cr10, respectively, compared with $39.5~\Omega^{-1}\,\mathrm{cm}^{-1}$ in MS. 
In particular, the optimal sample $\left[\mathrm{Mn}_{3}\mathrm{Sn}(2.4)/\mathrm{Cr}(0.15)\right]_{30}$ exhibits a room temperature AHC enhancement of about $46\%$ relative to undoped $\mathrm{Mn}_{3}\mathrm{Sn}$, in good agreement with the calculated low doping enhancement trend. 
When the temperature decreases to $50~\mathrm{K}$, it reaches $184.8$, $149.9$, and $109.2~\Omega^{-1}\,\mathrm{cm}^{-1}$ in $\left[\mathrm{Mn}_{3}\mathrm{Sn}(2.4)/\mathrm{Cr}(0.15)\right]_{30}$, $\left[\mathrm{Mn}_{3}\mathrm{Sn}(2.4)/\mathrm{Cr}(0.6)\right]_{30}$, and MS50Cr10, respectively, while $\sigma_{xy}$ in MS decreases to nearly zero. Additional $10~\mathrm{K}$ measurements on the modulation-doped samples show the same qualitative trend, with a substantially stronger Hall response retained relative to undoped MS. 
Additional $\Delta\rho_{xy}$ for samples with other Cr concentrations are provided in the Supplemental Material (Sec. SI14).
Thus, the Cr doped films achieve low temperature Hall retention comparable to other doped $\mathrm{Mn}_{3}\mathrm{Sn}$ systems, while simultaneously improving the room temperature AHC relative to undoped $\mathrm{Mn}_{3}\mathrm{Sn}$, in good agreement with the calculated trend of optimal doping at low concentrations.

\section{DISCUSSION}
In this work, first-principles calculations are first used to establish a concentration-dependent intrinsic picture of Cr-doped $\mathrm{Mn}_{3}\mathrm{Sn}$. Starting from the pristine compound, the AHC is associated with Berry curvature hotspots arising from specific band features near the Fermi level. VCA calculations identify a dilute Cr regime in which the intrinsic AHC is enhanced while the coplanar non-collinear AFM framework is largely retained. Detailed Berry curvature and Wannier Hamiltonian analyses reveal that Cr substitution amplifies the pre-existing Berry curvature hotspots through slight energy shifts, local gap narrowing, and redistribution of the relevant orbital weight. Explicit supercell calculations further complement this averaged picture by showing that the magnitude and tensor character of the AHC depend on the local Cr arrangement, while the low-Cr enhancement remains identifiable and the magnetic perturbation remains spatially localized.

The first-principles calculations establish the intrinsic Berry curvature contribution to the AHC in ideal periodic structures. In real sputtered films, grains with different orientations and their separating grain boundaries introduce additional structural and magnetic heterogeneity. Crystallographic texture, residual strain, structural defects, interfaces, compositional fluctuations, and impurity scattering further modify the local electronic and magnetic environments and introduce additional scattering-related contributions to the anomalous Hall response. Consequently, the measured Hall conductivity reflects both the intrinsic Berry curvature contribution and the structural and transport characteristics of the polycrystalline film and cannot be directly compared with the calculated intrinsic value. Nevertheless, a similar AHC enhancement tendency remains observable in both uniform and modulation-doped polycrystalline films. The differences in the magnitude and temperature dependence of the AHC between the two doping approaches further indicate that local Cr distribution, interfaces, and film microstructure can modulate the Hall response beyond the composition-averaged intrinsic picture. Taken together, the experiments show that the theoretically identified enhancement tendency persists under realistic thin-film conditions, while its detailed manifestation is further shaped by structural and disorder-related effects. A full treatment of configurational disorder and a quantitative separation of intrinsic and extrinsic Hall contributions would require disorder-aware transport calculations beyond the scope of the present work.

\begin{acknowledgments} 
L.S. thanks the financial support from Ministry of Education, Singapore under its Tier 2 Grants (No. A-8001872-00-00). YHW acknowledges the financial support from Ministry of Education, Singapore under its Tier 2 Grants (MOE2018-T2-1-076) and Advanced Research and Technology Innovation Centre (ARTIC), the National University of Singapore under Grant No. A-0005947-23-00. 
\end{acknowledgments}

\section{APPENDIX: METHODS}
\subsection{\label{sec:level6}DFT Calculations}
First-principles calculations were performed using \textsc{Quantum Espresso}~\cite{giannozziQUANTUMESPRESSOModular2009,giannozziAdvancedCapabilitiesMaterials2017}. The exchange-correlation interactions were treated within the generalized gradient approximation (GGA) using the Perdew-Burke-Ernzerhof (PBE) functional. Spin-orbit coupled (SOC) calculations were carried out self-consistently within the non-collinear formalism using fully relativistic ultrasoft pseudopotentials from PSLIBRARY~\cite{dalcorsoPseudopotentialsPeriodicTable2014}. For comparison, calculations without SOC were performed using scalar-relativistic pseudopotentials, which yielded a strictly zero anomalous Hall conductivity (AHC) in all subsequent calculations. Self-consistent field (SCF) calculations used an $8\times8\times8$ Monkhorst-Pack $k$-point mesh, with plane wave and charge density cutoffs of $60~\mathrm{Ry}$ and $480~\mathrm{Ry}$, respectively. 

Cr substitution on the Mn sublattice was modeled using both the virtual crystal approximation (VCA) and explicit supercells. In the VCA calculations, the spatially averaged Mn/Cr substitution was represented by a mixed fully relativistic ultrasoft pseudopotential generated in \textsc{Quantum Espresso} ~\cite{bellaicheVirtualCrystalApproximation2000} as
\begin{equation} 
V_{\mathrm{VCA}} = (1-\beta)V_{\mathrm{Mn}} + \beta V_{\mathrm{Cr}}, \label{eq:vca_potential} 
\end{equation}
where $\beta$ is the Mn-site Cr fraction defined in the main text. Using the same computational settings as for pristine $\mathrm{Mn}_{3}\mathrm{Sn}$, we considered VCA concentrations from $\beta=0\%$ to $25\%$. To resolve local configurational effects beyond the VCA description, five representative supercells were constructed by expanding the primitive cell through $c$-axis and in-plane expansion schemes, with one Mn atom replaced by Cr in each case, with the corresponding Mn site Cr fractions spanning $\beta=4.17\%$, $8.33\%$, and $16.7\%$. The $k$-point meshes were reduced along the expansion directions to maintain a reciprocal-space sampling density comparable to that of the primitive cell. Additional supercell details are provided in the Supplemental Material.

The Bloch states from the non-self-consistent field (NSCF) calculations were further transformed into maximally localized Wannier functions (MLWFs) using the WANNIER90 package~\cite{marzariMaximallyLocalizedWannier2012a}. In the Wannierization, $d$ orbitals were chosen as initial projections for all Mn-like sites, while $p$ orbitals were used for Sn. The AHC was then evaluated using WANNIERBERRI ~\cite{tsirkinHighPerformanceWannier2021} based on the Kubo formula~\cite{xiaoBerryPhaseEffects2010a,gradhandFirstprincipleCalculationsBerry2012a}:
\begin{equation} \sigma_{\alpha\beta} = -\frac{e^{2}}{\hbar} \sum_{n} \int_{\mathrm{BZ}} \frac{d^{3}\boldsymbol{k}}{(2\pi)^{3}} f_{n}(\boldsymbol{k}) \Omega_{n}^{\alpha\beta}(\boldsymbol{k}), \label{eq:ahc} \end{equation}
where $f_{n}(\boldsymbol{k})$ is the Fermi--Dirac distribution and $\Omega_{n}^{\alpha\beta}(\boldsymbol{k})$ is the band-resolved Berry curvature (BC),
\begin{equation} 
\Omega_{n}^{\alpha\beta}(\boldsymbol{k})
= 
-2\hbar^{2} 
\operatorname{Im} 
\sum_{m\neq n} 
\frac{ 
\langle u_{n\boldsymbol{k}} | 
\hat{v}_{\alpha} 
| u_{m\boldsymbol{k}} 
\rangle 
\langle u_{m\boldsymbol{k}} | \hat{v}_{\beta} | u_{n\boldsymbol{k}} \rangle }{ \left[ E_{n}(\boldsymbol{k})-E_{m}(\boldsymbol{k}) \right]^{2} }. 
\label{eq:bc} 
\end{equation}
 The velocity operator is defined here as $\hat{v}_{\alpha}=(i/\hbar)[\hat{H},\hat{r}_{\alpha}]$, and $\alpha$, $\beta$ and $\gamma$ represent the three Cartesian coordinates. The Brillouin zone integration was first performed on a $200\times200\times200$ $ k$-mesh and then further converged with 50 iterations of recursive adaptive refinement. Berry curvature hotspots were identified by scanning the three-dimensional Brillouin zone on a uniform $\boldsymbol{k}$-space grid using Wannier interpolation and locating regions with the largest absolute BC contributions. Representative $k_{z}$ planes and in-plane paths crossing these hotspots were then selected for BC mapping and band-resolved analysis at both the Fermi energy $E_{\mathrm{F}}$ and the energy corresponding to the maximum AHC $E_{\mathrm{max}}$.

\subsection{\label{sec:level7}Experiment}
\noindent\textbf{Thin-film growth and characterization.}
Two types of Cr doped $\mathrm{Mn}_{3}\mathrm{Sn}$ thin films were prepared using, respectively, uniform doping and modulation doping strategies, together with an undoped $\mathrm{Mn}_{3}\mathrm{Sn}$ reference. All films were deposited on $\mathrm{SiO}_{2}/\mathrm{Si}$ substrates by room temperature magnetron sputtering with a base pressure of $2 \times 10^{-8}~\mathrm{Torr}$ and a working pressure of $3 \times 10^{-3}~\mathrm{Torr}$, followed by post annealing at $450\,^{\circ}\mathrm{C}$ for $1~\mathrm{h}$ in vacuum ($2 \times 10^{-5}~\mathrm{Torr}$) to promote crystallization and dopant interdiffusion. Crystal structures were characterized by X-ray diffraction (XRD) and X-ray reflectivity (XRR) using a Rigaku SmartLab system with Cu K$\alpha$ radiation. Magnetic properties were measured using superconducting quantum interference device (SQUID) magnetometry.

\noindent\textbf{Electrical transport measurements.}
For anomalous Hall effect (AHE) and magnetoresistance (MR) measurements, the films were patterned into Hall bars with lateral dimensions of $150~\mu\mathrm{m} \times 15~\mu\mathrm{m}$ by photolithography, sputtering, and lift-off. Electrical transport measurements were performed using a Quantum Design VersaLab system. The Hall resistivity was described as

\begin{equation}
\rho_{xy}
=
R_{0}B
+
R_{\mathrm{s}}\mu_{0}M
+
\rho_{xy}^{\mathrm{AF}},
\label{eq:hall_resistivity}
\end{equation}
where $R_{0}$ and $R_{\mathrm{s}}$ are the ordinary and anomalous Hall coefficients, respectively, $B$ is the applied magnetic field, $\mu_{0}$ is the vacuum permeability, $M$ is the magnetization, and $\rho_{xy}^{\mathrm{AF}}$ denotes the Hall contribution associated with the inverse-triangular non-collinear antiferromagnetic state~\cite{khadkaKondoPhysicsAntiferromagnetic2020}. The zero-field Hall signal $\Delta\rho_{xy}$ discussed in the main text was extracted from the remanent response of the Hall loops.
\nocite{*}

\bibliography{draft}

\begin{thebibliography}{47}%
\makeatletter
\providecommand \@ifxundefined [1]{%
 \@ifx{#1\undefined}
}%
\providecommand \@ifnum [1]{%
 \ifnum #1\expandafter \@firstoftwo
 \else \expandafter \@secondoftwo
 \fi
}%
\providecommand \@ifx [1]{%
 \ifx #1\expandafter \@firstoftwo
 \else \expandafter \@secondoftwo
 \fi
}%
\providecommand \natexlab [1]{#1}%
\providecommand \enquote  [1]{``#1''}%
\providecommand \bibnamefont  [1]{#1}%
\providecommand \bibfnamefont [1]{#1}%
\providecommand \citenamefont [1]{#1}%
\providecommand \href@noop [0]{\@secondoftwo}%
\providecommand \href [0]{\begingroup \@sanitize@url \@href}%
\providecommand \@href[1]{\@@startlink{#1}\@@href}%
\providecommand \@@href[1]{\endgroup#1\@@endlink}%
\providecommand \@sanitize@url [0]{\catcode `\\12\catcode `\$12\catcode `\&12\catcode `\#12\catcode `\^12\catcode `\_12\catcode `\%12\relax}%
\providecommand \@@startlink[1]{}%
\providecommand \@@endlink[0]{}%
\providecommand \url  [0]{\begingroup\@sanitize@url \@url }%
\providecommand \@url [1]{\endgroup\@href {#1}{\urlprefix }}%
\providecommand \urlprefix  [0]{URL }%
\providecommand \Eprint [0]{\href }%
\providecommand \doibase [0]{https://doi.org/}%
\providecommand \selectlanguage [0]{\@gobble}%
\providecommand \bibinfo  [0]{\@secondoftwo}%
\providecommand \bibfield  [0]{\@secondoftwo}%
\providecommand \translation [1]{[#1]}%
\providecommand \BibitemOpen [0]{}%
\providecommand \bibitemStop [0]{}%
\providecommand \bibitemNoStop [0]{.\EOS\space}%
\providecommand \EOS [0]{\spacefactor3000\relax}%
\providecommand \BibitemShut  [1]{\csname bibitem#1\endcsname}%
\let\auto@bib@innerbib\@empty
\bibitem [{\citenamefont {Tsai}\ \emph {et~al.}(2026)\citenamefont {Tsai}, \citenamefont {Matsuda}, \citenamefont {Kondou}, \citenamefont {Shimizu}, \citenamefont {Nomoto}, \citenamefont {Higo}, \citenamefont {Matsuo}, \citenamefont {Tsushima}, \citenamefont {Asakura}, \citenamefont {Peng},\ and\ \citenamefont {{others}}}]{tsaiPicosecondUltralowpowerSwitching2026}%
  \BibitemOpen
  \bibfield  {author} {\bibinfo {author} {\bibfnamefont {H.}~\bibnamefont {Tsai}}, \bibinfo {author} {\bibfnamefont {T.}~\bibnamefont {Matsuda}}, \bibinfo {author} {\bibfnamefont {K.}~\bibnamefont {Kondou}}, \bibinfo {author} {\bibfnamefont {K.}~\bibnamefont {Shimizu}}, \bibinfo {author} {\bibfnamefont {T.}~\bibnamefont {Nomoto}}, \bibinfo {author} {\bibfnamefont {T.}~\bibnamefont {Higo}}, \bibinfo {author} {\bibfnamefont {T.}~\bibnamefont {Matsuo}}, \bibinfo {author} {\bibfnamefont {Y.}~\bibnamefont {Tsushima}}, \bibinfo {author} {\bibfnamefont {M.}~\bibnamefont {Asakura}}, \bibinfo {author} {\bibfnamefont {H.}~\bibnamefont {Peng}},\ and\ \bibinfo {author} {\bibnamefont {{others}}},\ }\bibfield  {title} {\bibinfo {title} {Picosecond ultralow-power switching device based on an antiferromagnet},\ }\href {https://doi.org/10.1126/science.adt3136} {\bibfield  {journal} {\bibinfo  {journal} {Science}\ }\textbf {\bibinfo {volume} {392}},\ \bibinfo {pages} {761} (\bibinfo {year} {2026})}\BibitemShut {NoStop}%
\bibitem [{\citenamefont {Zhou}\ \emph {et~al.}(2026)\citenamefont {Zhou}, \citenamefont {Cao}, \citenamefont {Pan}, \citenamefont {Zhang}, \citenamefont {Liang}, \citenamefont {Pan},\ and\ \citenamefont {Song}}]{zhouFieldfreeFullSwitching2026a}%
  \BibitemOpen
  \bibfield  {author} {\bibinfo {author} {\bibfnamefont {Z.}~\bibnamefont {Zhou}}, \bibinfo {author} {\bibfnamefont {Y.}~\bibnamefont {Cao}}, \bibinfo {author} {\bibfnamefont {Z.}~\bibnamefont {Pan}}, \bibinfo {author} {\bibfnamefont {Y.}~\bibnamefont {Zhang}}, \bibinfo {author} {\bibfnamefont {S.}~\bibnamefont {Liang}}, \bibinfo {author} {\bibfnamefont {F.}~\bibnamefont {Pan}},\ and\ \bibinfo {author} {\bibfnamefont {C.}~\bibnamefont {Song}},\ }\bibfield  {title} {\bibinfo {title} {Field-free full switching of chiral antiferromagnetic order},\ }\href {https://doi.org/10.1038/s41586-026-10175-6} {\bibfield  {journal} {\bibinfo  {journal} {Nature}\ }\textbf {\bibinfo {volume} {651}},\ \bibinfo {pages} {341} (\bibinfo {year} {2026})}\BibitemShut {NoStop}%
\bibitem [{\citenamefont {Rimmler}\ \emph {et~al.}(2024)\citenamefont {Rimmler}, \citenamefont {Pal},\ and\ \citenamefont {Parkin}}]{rimmlerNoncollinearAntiferromagneticSpintronics2025}%
  \BibitemOpen
  \bibfield  {author} {\bibinfo {author} {\bibfnamefont {B.~H.}\ \bibnamefont {Rimmler}}, \bibinfo {author} {\bibfnamefont {B.}~\bibnamefont {Pal}},\ and\ \bibinfo {author} {\bibfnamefont {S.~S.~P.}\ \bibnamefont {Parkin}},\ }\bibfield  {title} {\bibinfo {title} {Non-collinear antiferromagnetic spintronics},\ }\href {https://doi.org/10.1038/s41578-024-00706-w} {\bibfield  {journal} {\bibinfo  {journal} {Nature Reviews Materials}\ }\textbf {\bibinfo {volume} {10}},\ \bibinfo {pages} {109} (\bibinfo {year} {2024})}\BibitemShut {NoStop}%
\bibitem [{\citenamefont {Baltz}\ \emph {et~al.}(2018)\citenamefont {Baltz}, \citenamefont {Manchon}, \citenamefont {Tsoi}, \citenamefont {Moriyama}, \citenamefont {Ono},\ and\ \citenamefont {Tserkovnyak}}]{baltzAntiferromagneticSpintronics2018}%
  \BibitemOpen
  \bibfield  {author} {\bibinfo {author} {\bibfnamefont {V.}~\bibnamefont {Baltz}}, \bibinfo {author} {\bibfnamefont {A.}~\bibnamefont {Manchon}}, \bibinfo {author} {\bibfnamefont {M.}~\bibnamefont {Tsoi}}, \bibinfo {author} {\bibfnamefont {T.}~\bibnamefont {Moriyama}}, \bibinfo {author} {\bibfnamefont {T.}~\bibnamefont {Ono}},\ and\ \bibinfo {author} {\bibfnamefont {Y.}~\bibnamefont {Tserkovnyak}},\ }\bibfield  {title} {\bibinfo {title} {Antiferromagnetic spintronics},\ }\href {https://doi.org/10.1103/RevModPhys.90.015005} {\bibfield  {journal} {\bibinfo  {journal} {Reviews of Modern Physics}\ }\textbf {\bibinfo {volume} {90}},\ \bibinfo {pages} {015005} (\bibinfo {year} {2018})}\BibitemShut {NoStop}%
\bibitem [{\citenamefont {Dal~Din}\ \emph {et~al.}(2024)\citenamefont {Dal~Din}, \citenamefont {Amin}, \citenamefont {Wadley},\ and\ \citenamefont {Edmonds}}]{daldinAntiferromagneticSpintronics2024a}%
  \BibitemOpen
  \bibfield  {author} {\bibinfo {author} {\bibfnamefont {A.}~\bibnamefont {Dal~Din}}, \bibinfo {author} {\bibfnamefont {O.~J.}\ \bibnamefont {Amin}}, \bibinfo {author} {\bibfnamefont {P.}~\bibnamefont {Wadley}},\ and\ \bibinfo {author} {\bibfnamefont {K.~W.}\ \bibnamefont {Edmonds}},\ }\bibfield  {title} {\bibinfo {title} {Antiferromagnetic spintronics and beyond},\ }\href {https://doi.org/10.1038/s44306-024-00029-0} {\bibfield  {journal} {\bibinfo  {journal} {npj Spintronics}\ }\textbf {\bibinfo {volume} {2}},\ \bibinfo {pages} {25} (\bibinfo {year} {2024})}\BibitemShut {NoStop}%
\bibitem [{\citenamefont {Guo}\ and\ \citenamefont {Wang}(2017)}]{guoLargeAnomalousNernst2017}%
  \BibitemOpen
  \bibfield  {author} {\bibinfo {author} {\bibfnamefont {G.-Y.}\ \bibnamefont {Guo}}\ and\ \bibinfo {author} {\bibfnamefont {T.-C.}\ \bibnamefont {Wang}},\ }\bibfield  {title} {\bibinfo {title} {Large anomalous {{Nernst}} and spin {{Nernst}} effects in the noncollinear antiferromagnets {{Mn}} 3 {{X}} ( {{X}} = {{Sn}} , {{Ge}} , {{Ga}} )},\ }\href {https://doi.org/10.1103/PhysRevB.96.224415} {\bibfield  {journal} {\bibinfo  {journal} {Physical Review B}\ }\textbf {\bibinfo {volume} {96}},\ \bibinfo {pages} {224415} (\bibinfo {year} {2017})}\BibitemShut {NoStop}%
\bibitem [{\citenamefont {Chen}\ \emph {et~al.}(2014)\citenamefont {Chen}, \citenamefont {Niu},\ and\ \citenamefont {MacDonald}}]{chenAnomalousHallEffect2014a}%
  \BibitemOpen
  \bibfield  {author} {\bibinfo {author} {\bibfnamefont {H.}~\bibnamefont {Chen}}, \bibinfo {author} {\bibfnamefont {Q.}~\bibnamefont {Niu}},\ and\ \bibinfo {author} {\bibfnamefont {A.~H.}\ \bibnamefont {MacDonald}},\ }\bibfield  {title} {\bibinfo {title} {Anomalous {{Hall Effect Arising}} from {{Noncollinear Antiferromagnetism}}},\ }\href {https://doi.org/10.1103/PhysRevLett.112.017205} {\bibfield  {journal} {\bibinfo  {journal} {Physical Review Letters}\ }\textbf {\bibinfo {volume} {112}},\ \bibinfo {pages} {017205} (\bibinfo {year} {2014})}\BibitemShut {NoStop}%
\bibitem [{\citenamefont {Pan}\ \emph {et~al.}(2022)\citenamefont {Pan}, \citenamefont {Le}, \citenamefont {He}, \citenamefont {Watzman}, \citenamefont {Yao}, \citenamefont {Gooth}, \citenamefont {Heremans}, \citenamefont {Sun},\ and\ \citenamefont {Felser}}]{panGiantAnomalousNernst2022}%
  \BibitemOpen
  \bibfield  {author} {\bibinfo {author} {\bibfnamefont {Y.}~\bibnamefont {Pan}}, \bibinfo {author} {\bibfnamefont {C.}~\bibnamefont {Le}}, \bibinfo {author} {\bibfnamefont {B.}~\bibnamefont {He}}, \bibinfo {author} {\bibfnamefont {S.~J.}\ \bibnamefont {Watzman}}, \bibinfo {author} {\bibfnamefont {M.}~\bibnamefont {Yao}}, \bibinfo {author} {\bibfnamefont {J.}~\bibnamefont {Gooth}}, \bibinfo {author} {\bibfnamefont {J.~P.}\ \bibnamefont {Heremans}}, \bibinfo {author} {\bibfnamefont {Y.}~\bibnamefont {Sun}},\ and\ \bibinfo {author} {\bibfnamefont {C.}~\bibnamefont {Felser}},\ }\bibfield  {title} {\bibinfo {title} {Giant anomalous {{Nernst}} signal in the antiferromagnet {{YbMnBi2}}},\ }\href {https://doi.org/10.1038/s41563-021-01149-2} {\bibfield  {journal} {\bibinfo  {journal} {Nature Materials}\ }\textbf {\bibinfo {volume} {21}},\ \bibinfo {pages} {203} (\bibinfo {year} {2022})}\BibitemShut {NoStop}%
\bibitem [{\citenamefont {Reichlova}\ \emph {et~al.}(2019)\citenamefont {Reichlova}, \citenamefont {Janda}, \citenamefont {Godinho}, \citenamefont {Markou}, \citenamefont {Kriegner}, \citenamefont {Schlitz}, \citenamefont {Zelezny}, \citenamefont {Soban}, \citenamefont {Bejarano}, \citenamefont {Schultheiss} \emph {et~al.}}]{reichlovaImagingWritingMagnetic2019}%
  \BibitemOpen
  \bibfield  {author} {\bibinfo {author} {\bibfnamefont {H.}~\bibnamefont {Reichlova}}, \bibinfo {author} {\bibfnamefont {T.}~\bibnamefont {Janda}}, \bibinfo {author} {\bibfnamefont {J.}~\bibnamefont {Godinho}}, \bibinfo {author} {\bibfnamefont {A.}~\bibnamefont {Markou}}, \bibinfo {author} {\bibfnamefont {D.}~\bibnamefont {Kriegner}}, \bibinfo {author} {\bibfnamefont {R.}~\bibnamefont {Schlitz}}, \bibinfo {author} {\bibfnamefont {J.}~\bibnamefont {Zelezny}}, \bibinfo {author} {\bibfnamefont {Z.}~\bibnamefont {Soban}}, \bibinfo {author} {\bibfnamefont {M.}~\bibnamefont {Bejarano}}, \bibinfo {author} {\bibfnamefont {H.}~\bibnamefont {Schultheiss}}, \emph {et~al.},\ }\bibfield  {title} {\bibinfo {title} {Imaging and writing magnetic domains in the non-collinear antiferromagnet {{Mn3Sn}}},\ }\href {https://doi.org/10.1038/s41467-019-13391-z} {\bibfield  {journal} {\bibinfo  {journal} {Nature Communications}\ }\textbf {\bibinfo {volume} {10}},\ \bibinfo {pages} {5459} (\bibinfo {year} {2019})}\BibitemShut
  {NoStop}%
\bibitem [{\citenamefont {Ikhlas}\ \emph {et~al.}(2017)\citenamefont {Ikhlas}, \citenamefont {Tomita}, \citenamefont {Koretsune}, \citenamefont {Suzuki}, \citenamefont {{Nishio-Hamane}}, \citenamefont {Arita}, \citenamefont {Otani},\ and\ \citenamefont {Nakatsuji}}]{ikhlasLargeAnomalousNernst2017}%
  \BibitemOpen
  \bibfield  {author} {\bibinfo {author} {\bibfnamefont {M.}~\bibnamefont {Ikhlas}}, \bibinfo {author} {\bibfnamefont {T.}~\bibnamefont {Tomita}}, \bibinfo {author} {\bibfnamefont {T.}~\bibnamefont {Koretsune}}, \bibinfo {author} {\bibfnamefont {M.-T.}\ \bibnamefont {Suzuki}}, \bibinfo {author} {\bibfnamefont {D.}~\bibnamefont {{Nishio-Hamane}}}, \bibinfo {author} {\bibfnamefont {R.}~\bibnamefont {Arita}}, \bibinfo {author} {\bibfnamefont {Y.}~\bibnamefont {Otani}},\ and\ \bibinfo {author} {\bibfnamefont {S.}~\bibnamefont {Nakatsuji}},\ }\bibfield  {title} {\bibinfo {title} {Large anomalous {{Nernst}} effect at room temperature in a chiral antiferromagnet},\ }\href {https://doi.org/10.1038/nphys4181} {\bibfield  {journal} {\bibinfo  {journal} {Nature Physics}\ }\textbf {\bibinfo {volume} {13}},\ \bibinfo {pages} {1085} (\bibinfo {year} {2017})}\BibitemShut {NoStop}%
\bibitem [{\citenamefont {Qin}\ \emph {et~al.}(2023)\citenamefont {Qin}, \citenamefont {Yan}, \citenamefont {Wang}, \citenamefont {Chen}, \citenamefont {Meng}, \citenamefont {Dong}, \citenamefont {Zhu}, \citenamefont {Cai}, \citenamefont {Feng}, \citenamefont {Zhou} \emph {et~al.}}]{qinRoomtemperatureMagnetoresistanceAllantiferromagnetic2023}%
  \BibitemOpen
  \bibfield  {author} {\bibinfo {author} {\bibfnamefont {P.}~\bibnamefont {Qin}}, \bibinfo {author} {\bibfnamefont {H.}~\bibnamefont {Yan}}, \bibinfo {author} {\bibfnamefont {X.}~\bibnamefont {Wang}}, \bibinfo {author} {\bibfnamefont {H.}~\bibnamefont {Chen}}, \bibinfo {author} {\bibfnamefont {Z.}~\bibnamefont {Meng}}, \bibinfo {author} {\bibfnamefont {J.}~\bibnamefont {Dong}}, \bibinfo {author} {\bibfnamefont {M.}~\bibnamefont {Zhu}}, \bibinfo {author} {\bibfnamefont {J.}~\bibnamefont {Cai}}, \bibinfo {author} {\bibfnamefont {Z.}~\bibnamefont {Feng}}, \bibinfo {author} {\bibfnamefont {X.}~\bibnamefont {Zhou}}, \emph {et~al.},\ }\bibfield  {title} {\bibinfo {title} {Room-temperature magnetoresistance in an all-antiferromagnetic tunnel junction},\ }\href {https://doi.org/10.1038/s41586-022-05461-y} {\bibfield  {journal} {\bibinfo  {journal} {Nature}\ }\textbf {\bibinfo {volume} {613}},\ \bibinfo {pages} {485} (\bibinfo {year} {2023})}\BibitemShut {NoStop}%
\bibitem [{\citenamefont {Nagaosa}\ \emph {et~al.}(2010)\citenamefont {Nagaosa}, \citenamefont {Sinova}, \citenamefont {Onoda}, \citenamefont {MacDonald},\ and\ \citenamefont {Ong}}]{nagaosaAnomalousHallEffect2010a}%
  \BibitemOpen
  \bibfield  {author} {\bibinfo {author} {\bibfnamefont {N.}~\bibnamefont {Nagaosa}}, \bibinfo {author} {\bibfnamefont {J.}~\bibnamefont {Sinova}}, \bibinfo {author} {\bibfnamefont {S.}~\bibnamefont {Onoda}}, \bibinfo {author} {\bibfnamefont {A.~H.}\ \bibnamefont {MacDonald}},\ and\ \bibinfo {author} {\bibfnamefont {N.~P.}\ \bibnamefont {Ong}},\ }\bibfield  {title} {\bibinfo {title} {Anomalous {{Hall}} effect},\ }\href {https://doi.org/10.1103/RevModPhys.82.1539} {\bibfield  {journal} {\bibinfo  {journal} {Reviews of Modern Physics}\ }\textbf {\bibinfo {volume} {82}},\ \bibinfo {pages} {1539} (\bibinfo {year} {2010})}\BibitemShut {NoStop}%
\bibitem [{\citenamefont {Goren}\ \emph {et~al.}(2025)\citenamefont {Goren}, \citenamefont {Hazra}, \citenamefont {Zhang}, \citenamefont {Yochelis}, \citenamefont {Sun}, \citenamefont {Parkin},\ and\ \citenamefont {Paltiel}}]{gorenAntiferromagneticLocalVolatile2025}%
  \BibitemOpen
  \bibfield  {author} {\bibinfo {author} {\bibfnamefont {N.}~\bibnamefont {Goren}}, \bibinfo {author} {\bibfnamefont {B.~K.}\ \bibnamefont {Hazra}}, \bibinfo {author} {\bibfnamefont {T.-Y.}\ \bibnamefont {Zhang}}, \bibinfo {author} {\bibfnamefont {S.}~\bibnamefont {Yochelis}}, \bibinfo {author} {\bibfnamefont {Q.-F.}\ \bibnamefont {Sun}}, \bibinfo {author} {\bibfnamefont {S.~S.~P.}\ \bibnamefont {Parkin}},\ and\ \bibinfo {author} {\bibfnamefont {Y.}~\bibnamefont {Paltiel}},\ }\bibfield  {title} {\bibinfo {title} {Antiferromagnetic local volatile memory utilizing non-collinear {{Mn3Sn}} thin films and chiral gating},\ }\href {https://doi.org/10.1038/s43246-025-00954-5} {\bibfield  {journal} {\bibinfo  {journal} {Communications Materials}\ }\textbf {\bibinfo {volume} {6}},\ \bibinfo {pages} {238} (\bibinfo {year} {2025})}\BibitemShut {NoStop}%
\bibitem [{\citenamefont {Chen}\ \emph {et~al.}(2021{\natexlab{a}})\citenamefont {Chen}, \citenamefont {Tomita}, \citenamefont {Minami}, \citenamefont {Fu}, \citenamefont {Koretsune}, \citenamefont {Kitatani}, \citenamefont {Muhammad}, \citenamefont {{Nishio-Hamane}}, \citenamefont {Ishii}, \citenamefont {Ishii} \emph {et~al.}}]{chenAnomalousTransportDue2021}%
  \BibitemOpen
  \bibfield  {author} {\bibinfo {author} {\bibfnamefont {T.}~\bibnamefont {Chen}}, \bibinfo {author} {\bibfnamefont {T.}~\bibnamefont {Tomita}}, \bibinfo {author} {\bibfnamefont {S.}~\bibnamefont {Minami}}, \bibinfo {author} {\bibfnamefont {M.}~\bibnamefont {Fu}}, \bibinfo {author} {\bibfnamefont {T.}~\bibnamefont {Koretsune}}, \bibinfo {author} {\bibfnamefont {M.}~\bibnamefont {Kitatani}}, \bibinfo {author} {\bibfnamefont {I.}~\bibnamefont {Muhammad}}, \bibinfo {author} {\bibfnamefont {D.}~\bibnamefont {{Nishio-Hamane}}}, \bibinfo {author} {\bibfnamefont {R.}~\bibnamefont {Ishii}}, \bibinfo {author} {\bibfnamefont {F.}~\bibnamefont {Ishii}}, \emph {et~al.},\ }\bibfield  {title} {\bibinfo {title} {Anomalous transport due to {{Weyl}} fermions in the chiral antiferromagnets {{Mn3X}}, {{X}} = {{Sn}}, {{Ge}}},\ }\href {https://doi.org/10.1038/s41467-020-20838-1} {\bibfield  {journal} {\bibinfo  {journal} {Nature Communications}\ }\textbf {\bibinfo {volume} {12}},\ \bibinfo {pages} {572} (\bibinfo {year}
  {2021}{\natexlab{a}})}\BibitemShut {NoStop}%
\bibitem [{\citenamefont {Sung}\ \emph {et~al.}(2018)\citenamefont {Sung}, \citenamefont {Ronning}, \citenamefont {Thompson},\ and\ \citenamefont {Bauer}}]{sungMagneticPhaseDependence2018}%
  \BibitemOpen
  \bibfield  {author} {\bibinfo {author} {\bibfnamefont {N.~H.}\ \bibnamefont {Sung}}, \bibinfo {author} {\bibfnamefont {F.}~\bibnamefont {Ronning}}, \bibinfo {author} {\bibfnamefont {J.~D.}\ \bibnamefont {Thompson}},\ and\ \bibinfo {author} {\bibfnamefont {E.~D.}\ \bibnamefont {Bauer}},\ }\bibfield  {title} {\bibinfo {title} {Magnetic phase dependence of the anomalous {{Hall}} effect in {{Mn3Sn}} single crystals},\ }\href {https://doi.org/10.1063/1.5021133} {\bibfield  {journal} {\bibinfo  {journal} {Applied Physics Letters}\ }\textbf {\bibinfo {volume} {112}},\ \bibinfo {pages} {132406} (\bibinfo {year} {2018})}\BibitemShut {NoStop}%
\bibitem [{\citenamefont {Xie}\ \emph {et~al.}(2022{\natexlab{a}})\citenamefont {Xie}, \citenamefont {Chen}, \citenamefont {Zhang}, \citenamefont {Mu}, \citenamefont {Zhang}, \citenamefont {Yan},\ and\ \citenamefont {Wu}}]{xieMagnetizationSwitchingPolycrystalline2022b}%
  \BibitemOpen
  \bibfield  {author} {\bibinfo {author} {\bibfnamefont {H.}~\bibnamefont {Xie}}, \bibinfo {author} {\bibfnamefont {X.}~\bibnamefont {Chen}}, \bibinfo {author} {\bibfnamefont {Q.}~\bibnamefont {Zhang}}, \bibinfo {author} {\bibfnamefont {Z.}~\bibnamefont {Mu}}, \bibinfo {author} {\bibfnamefont {X.}~\bibnamefont {Zhang}}, \bibinfo {author} {\bibfnamefont {B.}~\bibnamefont {Yan}},\ and\ \bibinfo {author} {\bibfnamefont {Y.}~\bibnamefont {Wu}},\ }\bibfield  {title} {\bibinfo {title} {Magnetization switching in polycrystalline {{Mn3Sn}} thin film induced by self-generated spin-polarized current},\ }\href {https://doi.org/10.1038/s41467-022-33345-2} {\bibfield  {journal} {\bibinfo  {journal} {Nature Communications}\ }\textbf {\bibinfo {volume} {13}},\ \bibinfo {pages} {5744} (\bibinfo {year} {2022}{\natexlab{a}})}\BibitemShut {NoStop}%
\bibitem [{\citenamefont {Taylor}\ \emph {et~al.}(2020)\citenamefont {Taylor}, \citenamefont {Markou}, \citenamefont {Lesne}, \citenamefont {Sivakumar}, \citenamefont {Luo}, \citenamefont {Radu}, \citenamefont {Werner}, \citenamefont {Felser},\ and\ \citenamefont {Parkin}}]{taylorAnomalousTopologicalHall2020}%
  \BibitemOpen
  \bibfield  {author} {\bibinfo {author} {\bibfnamefont {J.~M.}\ \bibnamefont {Taylor}}, \bibinfo {author} {\bibfnamefont {A.}~\bibnamefont {Markou}}, \bibinfo {author} {\bibfnamefont {E.}~\bibnamefont {Lesne}}, \bibinfo {author} {\bibfnamefont {P.~K.}\ \bibnamefont {Sivakumar}}, \bibinfo {author} {\bibfnamefont {C.}~\bibnamefont {Luo}}, \bibinfo {author} {\bibfnamefont {F.}~\bibnamefont {Radu}}, \bibinfo {author} {\bibfnamefont {P.}~\bibnamefont {Werner}}, \bibinfo {author} {\bibfnamefont {C.}~\bibnamefont {Felser}},\ and\ \bibinfo {author} {\bibfnamefont {S.~S.~P.}\ \bibnamefont {Parkin}},\ }\bibfield  {title} {\bibinfo {title} {Anomalous and topological {{Hall}} effects in epitaxial thin films of the noncollinear antiferromagnet {{Mn}} 3 {{Sn}}},\ }\href {https://doi.org/10.1103/PhysRevB.101.094404} {\bibfield  {journal} {\bibinfo  {journal} {Physical Review B}\ }\textbf {\bibinfo {volume} {101}},\ \bibinfo {pages} {094404} (\bibinfo {year} {2020})}\BibitemShut {NoStop}%
\bibitem [{\citenamefont {Gas}\ \emph {et~al.}(2025)\citenamefont {Gas}, \citenamefont {Yoon}, \citenamefont {Sato}, \citenamefont {Kubota}, \citenamefont {D{\l}u{\.z}ewski}, \citenamefont {Kret}, \citenamefont {Domagala}, \citenamefont {Edathumkandy}, \citenamefont {Takeuchi}, \citenamefont {Kanai},\ and\ \citenamefont {{others}}}]{gasMagneticPhaseDiagram2025}%
  \BibitemOpen
  \bibfield  {author} {\bibinfo {author} {\bibfnamefont {K.}~\bibnamefont {Gas}}, \bibinfo {author} {\bibfnamefont {J.-Y.}\ \bibnamefont {Yoon}}, \bibinfo {author} {\bibfnamefont {Y.}~\bibnamefont {Sato}}, \bibinfo {author} {\bibfnamefont {H.}~\bibnamefont {Kubota}}, \bibinfo {author} {\bibfnamefont {P.}~\bibnamefont {D{\l}u{\.z}ewski}}, \bibinfo {author} {\bibfnamefont {S.}~\bibnamefont {Kret}}, \bibinfo {author} {\bibfnamefont {J.~Z.}\ \bibnamefont {Domagala}}, \bibinfo {author} {\bibfnamefont {Y.~K.}\ \bibnamefont {Edathumkandy}}, \bibinfo {author} {\bibfnamefont {Y.}~\bibnamefont {Takeuchi}}, \bibinfo {author} {\bibfnamefont {S.}~\bibnamefont {Kanai}},\ and\ \bibinfo {author} {\bibnamefont {{others}}},\ }\bibfield  {title} {\bibinfo {title} {Magnetic phase diagram of {{Mn3}}+{{xSn1}}-x epitaxial thin films: {{Extending}} the anomalous hall effect to low temperatures via intrinsic alloying},\ }\href {https://doi.org/10.1063/5.0254918} {\bibfield  {journal} {\bibinfo  {journal} {APL Materials}\ }\textbf
  {\bibinfo {volume} {13}},\ \bibinfo {pages} {041105} (\bibinfo {year} {2025})}\BibitemShut {NoStop}%
\bibitem [{\citenamefont {K{\"u}bler}\ and\ \citenamefont {Felser}(2017)}]{kublerWeylFermionsAntiferromagnetic2018}%
  \BibitemOpen
  \bibfield  {author} {\bibinfo {author} {\bibfnamefont {J.}~\bibnamefont {K{\"u}bler}}\ and\ \bibinfo {author} {\bibfnamefont {C.}~\bibnamefont {Felser}},\ }\bibfield  {title} {\bibinfo {title} {Weyl fermions in antiferromagnetic {{Mn}}{\textsubscript{3}} {{Sn}} and {{Mn}}{\textsubscript{3}} {{Ge}}},\ }\href {https://doi.org/10.1209/0295-5075/120/47002} {\bibfield  {journal} {\bibinfo  {journal} {EPL (Europhysics Letters)}\ }\textbf {\bibinfo {volume} {120}},\ \bibinfo {pages} {47002} (\bibinfo {year} {2017})}\BibitemShut {NoStop}%
\bibitem [{\citenamefont {Khadka}\ \emph {et~al.}(2020)\citenamefont {Khadka}, \citenamefont {Thapaliya}, \citenamefont {Hurtado~Parra}, \citenamefont {Han}, \citenamefont {Wen}, \citenamefont {Need}, \citenamefont {Khanal}, \citenamefont {Wang}, \citenamefont {Zang}, \citenamefont {Kikkawa} \emph {et~al.}}]{khadkaKondoPhysicsAntiferromagnetic2020}%
  \BibitemOpen
  \bibfield  {author} {\bibinfo {author} {\bibfnamefont {D.}~\bibnamefont {Khadka}}, \bibinfo {author} {\bibfnamefont {T.~R.}\ \bibnamefont {Thapaliya}}, \bibinfo {author} {\bibfnamefont {S.}~\bibnamefont {Hurtado~Parra}}, \bibinfo {author} {\bibfnamefont {X.}~\bibnamefont {Han}}, \bibinfo {author} {\bibfnamefont {J.}~\bibnamefont {Wen}}, \bibinfo {author} {\bibfnamefont {R.~F.}\ \bibnamefont {Need}}, \bibinfo {author} {\bibfnamefont {P.}~\bibnamefont {Khanal}}, \bibinfo {author} {\bibfnamefont {W.}~\bibnamefont {Wang}}, \bibinfo {author} {\bibfnamefont {J.}~\bibnamefont {Zang}}, \bibinfo {author} {\bibfnamefont {J.~M.}\ \bibnamefont {Kikkawa}}, \emph {et~al.},\ }\bibfield  {title} {\bibinfo {title} {Kondo physics in antiferromagnetic {{Weyl}} semimetal {{Mn}}{\textsubscript{ 3+ {\emph{x}} }} {{Sn}}{\textsubscript{ 1- {\emph{x}} }} films},\ }\href {https://doi.org/10.1126/sciadv.abc1977} {\bibfield  {journal} {\bibinfo  {journal} {Science Advances}\ }\textbf {\bibinfo {volume} {6}},\ \bibinfo {pages}
  {eabc1977} (\bibinfo {year} {2020})}\BibitemShut {NoStop}%
\bibitem [{\citenamefont {Bangar}\ \emph {et~al.}(2023)\citenamefont {Bangar}, \citenamefont {Khan}, \citenamefont {Kumar}, \citenamefont {Chowdhury}, \citenamefont {Muduli},\ and\ \citenamefont {Muduli}}]{bangarLargeSpinHall2023}%
  \BibitemOpen
  \bibfield  {author} {\bibinfo {author} {\bibfnamefont {H.}~\bibnamefont {Bangar}}, \bibinfo {author} {\bibfnamefont {K.~I.~A.}\ \bibnamefont {Khan}}, \bibinfo {author} {\bibfnamefont {A.}~\bibnamefont {Kumar}}, \bibinfo {author} {\bibfnamefont {N.}~\bibnamefont {Chowdhury}}, \bibinfo {author} {\bibfnamefont {P.~K.}\ \bibnamefont {Muduli}},\ and\ \bibinfo {author} {\bibfnamefont {P.~K.}\ \bibnamefont {Muduli}},\ }\bibfield  {title} {\bibinfo {title} {Large {{Spin Hall Conductivity}} in {{Epitaxial Thin Films}} of {{Kagome Antiferromagnet Mn}}{\textsubscript{3}} {{Sn}} at {{Room Temperature}}},\ }\href {https://doi.org/10.1002/qute.202200115} {\bibfield  {journal} {\bibinfo  {journal} {Advanced Quantum Technologies}\ }\textbf {\bibinfo {volume} {6}},\ \bibinfo {pages} {2200115} (\bibinfo {year} {2023})}\BibitemShut {NoStop}%
\bibitem [{\citenamefont {Wen}\ \emph {et~al.}(2025)\citenamefont {Wen}, \citenamefont {Xiong}, \citenamefont {Yang}, \citenamefont {Zhu},\ and\ \citenamefont {Zhao}}]{wenEnhancedAnomalousHall2025}%
  \BibitemOpen
  \bibfield  {author} {\bibinfo {author} {\bibfnamefont {C.}~\bibnamefont {Wen}}, \bibinfo {author} {\bibfnamefont {D.}~\bibnamefont {Xiong}}, \bibinfo {author} {\bibfnamefont {C.}~\bibnamefont {Yang}}, \bibinfo {author} {\bibfnamefont {D.}~\bibnamefont {Zhu}},\ and\ \bibinfo {author} {\bibfnamefont {W.}~\bibnamefont {Zhao}},\ }\bibfield  {title} {\bibinfo {title} {Enhanced anomalous {{Hall}} conductivity via {{Ga}} doping in {{Mn}} 3 {{Sn}} and {{Mn}} 3 {{Ge}}},\ }\href {https://doi.org/10.1103/tktp-dhvc} {\bibfield  {journal} {\bibinfo  {journal} {Physical Review B}\ }\textbf {\bibinfo {volume} {112}},\ \bibinfo {pages} {115118} (\bibinfo {year} {2025})}\BibitemShut {NoStop}%
\bibitem [{\citenamefont {Raju}\ \emph {et~al.}(2025)\citenamefont {Raju}, \citenamefont {Matsuo}, \citenamefont {Balkew}, \citenamefont {{Nishio-Hamane}}, \citenamefont {Higo}, \citenamefont {Broholm},\ and\ \citenamefont {Nakatsuji}}]{rajuTunableHighNeel2025a}%
  \BibitemOpen
  \bibfield  {author} {\bibinfo {author} {\bibfnamefont {M.}~\bibnamefont {Raju}}, \bibinfo {author} {\bibfnamefont {T.}~\bibnamefont {Matsuo}}, \bibinfo {author} {\bibfnamefont {M.~B.}\ \bibnamefont {Balkew}}, \bibinfo {author} {\bibfnamefont {D.}~\bibnamefont {{Nishio-Hamane}}}, \bibinfo {author} {\bibfnamefont {T.}~\bibnamefont {Higo}}, \bibinfo {author} {\bibfnamefont {C.}~\bibnamefont {Broholm}},\ and\ \bibinfo {author} {\bibfnamefont {S.}~\bibnamefont {Nakatsuji}},\ }\bibfield  {title} {\bibinfo {title} {Tunable high {{N\'eel}} temperature and large anomalous {{Hall}} response in antiferromagnetic {{Weyl}} semimetal {{Mn3Sn1}}-{{xGax}} thin films},\ }\href {https://doi.org/10.1038/s43246-025-00912-1} {\bibfield  {journal} {\bibinfo  {journal} {Communications Materials}\ }\textbf {\bibinfo {volume} {6}},\ \bibinfo {pages} {205} (\bibinfo {year} {2025})}\BibitemShut {NoStop}%
\bibitem [{\citenamefont {Yano}\ \emph {et~al.}(2024)\citenamefont {Yano}, \citenamefont {Kabeya}, \citenamefont {Hayashi}, \citenamefont {Kihara}, \citenamefont {Suto}, \citenamefont {Yoshida}, \citenamefont {Yamaguchi},\ and\ \citenamefont {Kashiwaya}}]{yanoTuningAnomalousHall2024b}%
  \BibitemOpen
  \bibfield  {author} {\bibinfo {author} {\bibfnamefont {R.}~\bibnamefont {Yano}}, \bibinfo {author} {\bibfnamefont {T.}~\bibnamefont {Kabeya}}, \bibinfo {author} {\bibfnamefont {M.}~\bibnamefont {Hayashi}}, \bibinfo {author} {\bibfnamefont {S.}~\bibnamefont {Kihara}}, \bibinfo {author} {\bibfnamefont {H.}~\bibnamefont {Suto}}, \bibinfo {author} {\bibfnamefont {K.}~\bibnamefont {Yoshida}}, \bibinfo {author} {\bibfnamefont {T.}~\bibnamefont {Yamaguchi}},\ and\ \bibinfo {author} {\bibfnamefont {S.}~\bibnamefont {Kashiwaya}},\ }\bibfield  {title} {\bibinfo {title} {Tuning of {{Anomalous Hall Effect}} by {{Systematic Doping}} on {{Mn}}{\textsubscript{3}} {{Sn}}},\ }\href {https://doi.org/10.1021/acs.cgd.4c00104} {\bibfield  {journal} {\bibinfo  {journal} {Crystal Growth \& Design}\ }\textbf {\bibinfo {volume} {24}},\ \bibinfo {pages} {6124} (\bibinfo {year} {2024})}\BibitemShut {NoStop}%
\bibitem [{\citenamefont {Liu}\ \emph {et~al.}(2020)\citenamefont {Liu}, \citenamefont {Zuo}, \citenamefont {Li}, \citenamefont {Liu}, \citenamefont {Zheng}, \citenamefont {Zhang}, \citenamefont {Zhao}, \citenamefont {Hu}, \citenamefont {Sun},\ and\ \citenamefont {Shen}}]{liuSpontaneousMagneticBubbles2020}%
  \BibitemOpen
  \bibfield  {author} {\bibinfo {author} {\bibfnamefont {J.}~\bibnamefont {Liu}}, \bibinfo {author} {\bibfnamefont {S.}~\bibnamefont {Zuo}}, \bibinfo {author} {\bibfnamefont {H.}~\bibnamefont {Li}}, \bibinfo {author} {\bibfnamefont {Y.}~\bibnamefont {Liu}}, \bibinfo {author} {\bibfnamefont {X.}~\bibnamefont {Zheng}}, \bibinfo {author} {\bibfnamefont {Y.}~\bibnamefont {Zhang}}, \bibinfo {author} {\bibfnamefont {T.}~\bibnamefont {Zhao}}, \bibinfo {author} {\bibfnamefont {F.}~\bibnamefont {Hu}}, \bibinfo {author} {\bibfnamefont {J.}~\bibnamefont {Sun}},\ and\ \bibinfo {author} {\bibfnamefont {B.}~\bibnamefont {Shen}},\ }\bibfield  {title} {\bibinfo {title} {Spontaneous magnetic bubbles and large topological {{Hall}} effect in {{Mn3-xFexSn}} compound},\ }\href {https://doi.org/10.1016/j.scriptamat.2020.06.034} {\bibfield  {journal} {\bibinfo  {journal} {Scripta Materialia}\ }\textbf {\bibinfo {volume} {187}},\ \bibinfo {pages} {268} (\bibinfo {year} {2020})}\BibitemShut {NoStop}%
\bibitem [{\citenamefont {Sapkota}\ \emph {et~al.}(2025)\citenamefont {Sapkota}, \citenamefont {Llamazares}, \citenamefont {Hofer}, \citenamefont {Wetzel}, \citenamefont {Stiwinter}, \citenamefont {Griswold},\ and\ \citenamefont {Mazumdar}}]{sapkotaSynthesisPropertiesStoichiometric2025a}%
  \BibitemOpen
  \bibfield  {author} {\bibinfo {author} {\bibfnamefont {Y.~R.}\ \bibnamefont {Sapkota}}, \bibinfo {author} {\bibfnamefont {J.}~\bibnamefont {Llamazares}}, \bibinfo {author} {\bibfnamefont {S.}~\bibnamefont {Hofer}}, \bibinfo {author} {\bibfnamefont {D.}~\bibnamefont {Wetzel}}, \bibinfo {author} {\bibfnamefont {K.}~\bibnamefont {Stiwinter}}, \bibinfo {author} {\bibfnamefont {L.}~\bibnamefont {Griswold}},\ and\ \bibinfo {author} {\bibfnamefont {D.}~\bibnamefont {Mazumdar}},\ }\bibfield  {title} {\bibinfo {title} {Synthesis and properties of stoichiometric {{Mn3-xFexSn}} thin films},\ }\href {https://doi.org/10.1016/j.nxmate.2025.101211} {\bibfield  {journal} {\bibinfo  {journal} {Next Materials}\ }\textbf {\bibinfo {volume} {9}},\ \bibinfo {pages} {101211} (\bibinfo {year} {2025})}\BibitemShut {NoStop}%
\bibitem [{\citenamefont {Low}\ \emph {et~al.}(2022)\citenamefont {Low}, \citenamefont {Ghosh}, \citenamefont {Changdar}, \citenamefont {Routh}, \citenamefont {Purwar},\ and\ \citenamefont {Thirupathaiah}}]{lowTuningTopologicalProperties2022}%
  \BibitemOpen
  \bibfield  {author} {\bibinfo {author} {\bibfnamefont {A.}~\bibnamefont {Low}}, \bibinfo {author} {\bibfnamefont {S.}~\bibnamefont {Ghosh}}, \bibinfo {author} {\bibfnamefont {S.}~\bibnamefont {Changdar}}, \bibinfo {author} {\bibfnamefont {S.}~\bibnamefont {Routh}}, \bibinfo {author} {\bibfnamefont {S.}~\bibnamefont {Purwar}},\ and\ \bibinfo {author} {\bibfnamefont {S.}~\bibnamefont {Thirupathaiah}},\ }\bibfield  {title} {\bibinfo {title} {Tuning of topological properties in the strongly correlated antiferromagnet {{Mn}} 3 {{Sn}} via {{Fe}} doping},\ }\href {https://doi.org/10.1103/PhysRevB.106.144429} {\bibfield  {journal} {\bibinfo  {journal} {Physical Review B}\ }\textbf {\bibinfo {volume} {106}},\ \bibinfo {pages} {144429} (\bibinfo {year} {2022})}\BibitemShut {NoStop}%
\bibitem [{\citenamefont {Singh}\ \emph {et~al.}(2024)\citenamefont {Singh}, \citenamefont {Jamaluddin}, \citenamefont {Pradhan}, \citenamefont {Nandy}, \citenamefont {Tokunaga}, \citenamefont {Avdeev},\ and\ \citenamefont {Nayak}}]{singhHigherOrderExchange2024}%
  \BibitemOpen
  \bibfield  {author} {\bibinfo {author} {\bibfnamefont {C.}~\bibnamefont {Singh}}, \bibinfo {author} {\bibfnamefont {S.}~\bibnamefont {Jamaluddin}}, \bibinfo {author} {\bibfnamefont {S.}~\bibnamefont {Pradhan}}, \bibinfo {author} {\bibfnamefont {A.~K.}\ \bibnamefont {Nandy}}, \bibinfo {author} {\bibfnamefont {M.}~\bibnamefont {Tokunaga}}, \bibinfo {author} {\bibfnamefont {M.}~\bibnamefont {Avdeev}},\ and\ \bibinfo {author} {\bibfnamefont {A.~K.}\ \bibnamefont {Nayak}},\ }\bibfield  {title} {\bibinfo {title} {Higher order exchange driven noncoplanar magnetic state and large anomalous {{Hall}} effects in electron doped kagome magnet {{Mn3Sn}}},\ }\href {https://doi.org/10.1038/s41535-024-00657-z} {\bibfield  {journal} {\bibinfo  {journal} {npj Quantum Materials}\ }\textbf {\bibinfo {volume} {9}},\ \bibinfo {pages} {43} (\bibinfo {year} {2024})}\BibitemShut {NoStop}%
\bibitem [{\citenamefont {Chen}\ \emph {et~al.}(2021{\natexlab{b}})\citenamefont {Chen}, \citenamefont {Xie}, \citenamefont {Zhang}, \citenamefont {Luo}, \citenamefont {Shen},\ and\ \citenamefont {Wu}}]{chenGiantEnhancementAnomalous2021}%
  \BibitemOpen
  \bibfield  {author} {\bibinfo {author} {\bibfnamefont {X.}~\bibnamefont {Chen}}, \bibinfo {author} {\bibfnamefont {H.}~\bibnamefont {Xie}}, \bibinfo {author} {\bibfnamefont {Q.}~\bibnamefont {Zhang}}, \bibinfo {author} {\bibfnamefont {Z.}~\bibnamefont {Luo}}, \bibinfo {author} {\bibfnamefont {L.}~\bibnamefont {Shen}},\ and\ \bibinfo {author} {\bibfnamefont {Y.}~\bibnamefont {Wu}},\ }\href {https://doi.org/10.48550/arXiv.2107.00959} {\bibinfo {title} {Giant enhancement of anomalous hall effect in cr modulation-doped non-collinear antiferromagnetic {{Mn3Sn}} thin films}} (\bibinfo {year} {2021}{\natexlab{b}}),\ \Eprint {https://arxiv.org/abs/2107.00959} {arXiv:2107.00959} \BibitemShut {NoStop}%
\bibitem [{\citenamefont {Hori}\ \emph {et~al.}(1990)\citenamefont {Hori}, \citenamefont {Niida}, \citenamefont {Yamaguchi}, \citenamefont {Kato},\ and\ \citenamefont {Nakagawa}}]{horiAntiferromagneticFerromagneticTransition1990}%
  \BibitemOpen
  \bibfield  {author} {\bibinfo {author} {\bibfnamefont {T.}~\bibnamefont {Hori}}, \bibinfo {author} {\bibfnamefont {H.}~\bibnamefont {Niida}}, \bibinfo {author} {\bibfnamefont {Y.}~\bibnamefont {Yamaguchi}}, \bibinfo {author} {\bibfnamefont {H.}~\bibnamefont {Kato}},\ and\ \bibinfo {author} {\bibfnamefont {Y.}~\bibnamefont {Nakagawa}},\ }\bibfield  {title} {\bibinfo {title} {Antiferromagnetic to ferromagnetic transition of {{DO19}} type ({{Mn1-xFex}}){{3Sn1-$\delta$}}},\ }\href {https://doi.org/10.1016/S0304-8853(10)80053-8} {\bibfield  {journal} {\bibinfo  {journal} {Journal of Magnetism and Magnetic Materials}\ }\textbf {\bibinfo {volume} {90--91}},\ \bibinfo {pages} {159} (\bibinfo {year} {1990})}\BibitemShut {NoStop}%
\bibitem [{\citenamefont {Soh}\ \emph {et~al.}(2020)\citenamefont {Soh}, \citenamefont {De~Juan}, \citenamefont {Qureshi}, \citenamefont {Jacobsen}, \citenamefont {Wang}, \citenamefont {Guo},\ and\ \citenamefont {Boothroyd}}]{sohGroundstateMagneticStructure2020}%
  \BibitemOpen
  \bibfield  {author} {\bibinfo {author} {\bibfnamefont {J.-R.}\ \bibnamefont {Soh}}, \bibinfo {author} {\bibfnamefont {F.}~\bibnamefont {De~Juan}}, \bibinfo {author} {\bibfnamefont {N.}~\bibnamefont {Qureshi}}, \bibinfo {author} {\bibfnamefont {H.}~\bibnamefont {Jacobsen}}, \bibinfo {author} {\bibfnamefont {H.-Y.}\ \bibnamefont {Wang}}, \bibinfo {author} {\bibfnamefont {Y.-F.}\ \bibnamefont {Guo}},\ and\ \bibinfo {author} {\bibfnamefont {A.~T.}\ \bibnamefont {Boothroyd}},\ }\bibfield  {title} {\bibinfo {title} {Ground-state magnetic structure of {{Mn}} 3 {{Ge}}},\ }\href {https://doi.org/10.1103/PhysRevB.101.140411} {\bibfield  {journal} {\bibinfo  {journal} {Physical Review B}\ }\textbf {\bibinfo {volume} {101}},\ \bibinfo {pages} {140411} (\bibinfo {year} {2020})}\BibitemShut {NoStop}%
\bibitem [{SM()}]{SM}%
  \BibitemOpen
  \href@noop {} {}\bibinfo {note} {See Supplemental Material at [URL will be inserted by publisher] for additional computational and experimental details, which includes Refs.~\cite{ikedaFabricationEvaluationHighly2020,xieMagnetizationSwitchingPolycrystalline2022a}}\BibitemShut {NoStop}%
\bibitem [{\citenamefont {Xiao}\ \emph {et~al.}(2010)\citenamefont {Xiao}, \citenamefont {Chang},\ and\ \citenamefont {Niu}}]{xiaoBerryPhaseEffects2010a}%
  \BibitemOpen
  \bibfield  {author} {\bibinfo {author} {\bibfnamefont {D.}~\bibnamefont {Xiao}}, \bibinfo {author} {\bibfnamefont {M.-C.}\ \bibnamefont {Chang}},\ and\ \bibinfo {author} {\bibfnamefont {Q.}~\bibnamefont {Niu}},\ }\bibfield  {title} {\bibinfo {title} {Berry phase effects on electronic properties},\ }\href {https://doi.org/10.1103/RevModPhys.82.1959} {\bibfield  {journal} {\bibinfo  {journal} {Reviews of Modern Physics}\ }\textbf {\bibinfo {volume} {82}},\ \bibinfo {pages} {1959} (\bibinfo {year} {2010})}\BibitemShut {NoStop}%
\bibitem [{\citenamefont {Zhao}\ \emph {et~al.}(2021)\citenamefont {Zhao}, \citenamefont {Guo}, \citenamefont {Chen}, \citenamefont {Zhang},\ and\ \citenamefont {Jiang}}]{zhaoMagneticPropertiesAnomalous2021}%
  \BibitemOpen
  \bibfield  {author} {\bibinfo {author} {\bibfnamefont {Z.}~\bibnamefont {Zhao}}, \bibinfo {author} {\bibfnamefont {Q.}~\bibnamefont {Guo}}, \bibinfo {author} {\bibfnamefont {F.}~\bibnamefont {Chen}}, \bibinfo {author} {\bibfnamefont {K.}~\bibnamefont {Zhang}},\ and\ \bibinfo {author} {\bibfnamefont {Y.}~\bibnamefont {Jiang}},\ }\bibfield  {title} {\bibinfo {title} {Magnetic properties and anomalous {{Hall}} effect in antiferromagnetic {{Mn3Sn}} films},\ }\href {https://doi.org/10.1016/j.physb.2020.412692} {\bibfield  {journal} {\bibinfo  {journal} {Physica B: Condensed Matter}\ }\textbf {\bibinfo {volume} {604}},\ \bibinfo {pages} {412692} (\bibinfo {year} {2021})}\BibitemShut {NoStop}%
\bibitem [{\citenamefont {Wang}\ \emph {et~al.}(2019)\citenamefont {Wang}, \citenamefont {Feng}, \citenamefont {Qin}, \citenamefont {Yan}, \citenamefont {Zhou}, \citenamefont {Guo}, \citenamefont {Leng}, \citenamefont {Chen}, \citenamefont {Jia}, \citenamefont {Hu} \emph {et~al.}}]{wangIntegrationNoncollinearAntiferromagnetic2019}%
  \BibitemOpen
  \bibfield  {author} {\bibinfo {author} {\bibfnamefont {X.}~\bibnamefont {Wang}}, \bibinfo {author} {\bibfnamefont {Z.}~\bibnamefont {Feng}}, \bibinfo {author} {\bibfnamefont {P.}~\bibnamefont {Qin}}, \bibinfo {author} {\bibfnamefont {H.}~\bibnamefont {Yan}}, \bibinfo {author} {\bibfnamefont {X.}~\bibnamefont {Zhou}}, \bibinfo {author} {\bibfnamefont {H.}~\bibnamefont {Guo}}, \bibinfo {author} {\bibfnamefont {Z.}~\bibnamefont {Leng}}, \bibinfo {author} {\bibfnamefont {W.}~\bibnamefont {Chen}}, \bibinfo {author} {\bibfnamefont {Q.}~\bibnamefont {Jia}}, \bibinfo {author} {\bibfnamefont {Z.}~\bibnamefont {Hu}}, \emph {et~al.},\ }\bibfield  {title} {\bibinfo {title} {Integration of the noncollinear antiferromagnetic metal {{Mn3Sn}} onto ferroelectric oxides for electric-field control},\ }\href {https://doi.org/10.1016/j.actamat.2019.10.020} {\bibfield  {journal} {\bibinfo  {journal} {Acta Materialia}\ }\textbf {\bibinfo {volume} {181}},\ \bibinfo {pages} {537} (\bibinfo {year} {2019})}\BibitemShut {NoStop}%
\bibitem [{\citenamefont {Markou}\ \emph {et~al.}(2018)\citenamefont {Markou}, \citenamefont {Taylor}, \citenamefont {Kalache}, \citenamefont {Werner}, \citenamefont {Parkin},\ and\ \citenamefont {Felser}}]{markouNoncollinearAntiferromagneticMn3Sn2018}%
  \BibitemOpen
  \bibfield  {author} {\bibinfo {author} {\bibfnamefont {A.}~\bibnamefont {Markou}}, \bibinfo {author} {\bibfnamefont {J.~M.}\ \bibnamefont {Taylor}}, \bibinfo {author} {\bibfnamefont {A.}~\bibnamefont {Kalache}}, \bibinfo {author} {\bibfnamefont {P.}~\bibnamefont {Werner}}, \bibinfo {author} {\bibfnamefont {S.~S.~P.}\ \bibnamefont {Parkin}},\ and\ \bibinfo {author} {\bibfnamefont {C.}~\bibnamefont {Felser}},\ }\bibfield  {title} {\bibinfo {title} {Noncollinear antiferromagnetic {{Mn}} 3 {{Sn}} films},\ }\href {https://doi.org/10.1103/PhysRevMaterials.2.051001} {\bibfield  {journal} {\bibinfo  {journal} {Physical Review Materials}\ }\textbf {\bibinfo {volume} {2}},\ \bibinfo {pages} {051001} (\bibinfo {year} {2018})}\BibitemShut {NoStop}%
\bibitem [{\citenamefont {Ikeda}\ \emph {et~al.}(2018)\citenamefont {Ikeda}, \citenamefont {Tsunoda}, \citenamefont {Oogane}, \citenamefont {Oh}, \citenamefont {Morita},\ and\ \citenamefont {Ando}}]{ikedaAnomalousHallEffect2018}%
  \BibitemOpen
  \bibfield  {author} {\bibinfo {author} {\bibfnamefont {T.}~\bibnamefont {Ikeda}}, \bibinfo {author} {\bibfnamefont {M.}~\bibnamefont {Tsunoda}}, \bibinfo {author} {\bibfnamefont {M.}~\bibnamefont {Oogane}}, \bibinfo {author} {\bibfnamefont {S.}~\bibnamefont {Oh}}, \bibinfo {author} {\bibfnamefont {T.}~\bibnamefont {Morita}},\ and\ \bibinfo {author} {\bibfnamefont {Y.}~\bibnamefont {Ando}},\ }\bibfield  {title} {\bibinfo {title} {Anomalous {{Hall}} effect in polycrystalline {{Mn3Sn}} thin films},\ }\href {https://doi.org/10.1063/1.5051495} {\bibfield  {journal} {\bibinfo  {journal} {Applied Physics Letters}\ }\textbf {\bibinfo {volume} {113}},\ \bibinfo {pages} {222405} (\bibinfo {year} {2018})}\BibitemShut {NoStop}%
\bibitem [{\citenamefont {Higo}\ \emph {et~al.}(2018)\citenamefont {Higo}, \citenamefont {Qu}, \citenamefont {Li}, \citenamefont {Chien}, \citenamefont {Otani},\ and\ \citenamefont {Nakatsuji}}]{higoAnomalousHallEffect2018}%
  \BibitemOpen
  \bibfield  {author} {\bibinfo {author} {\bibfnamefont {T.}~\bibnamefont {Higo}}, \bibinfo {author} {\bibfnamefont {D.}~\bibnamefont {Qu}}, \bibinfo {author} {\bibfnamefont {Y.}~\bibnamefont {Li}}, \bibinfo {author} {\bibfnamefont {C.~L.}\ \bibnamefont {Chien}}, \bibinfo {author} {\bibfnamefont {Y.}~\bibnamefont {Otani}},\ and\ \bibinfo {author} {\bibfnamefont {S.}~\bibnamefont {Nakatsuji}},\ }\bibfield  {title} {\bibinfo {title} {Anomalous {{Hall}} effect in thin films of the {{Weyl}} antiferromagnet {{Mn3Sn}}},\ }\href {https://doi.org/10.1063/1.5064697} {\bibfield  {journal} {\bibinfo  {journal} {Applied Physics Letters}\ }\textbf {\bibinfo {volume} {113}},\ \bibinfo {pages} {202402} (\bibinfo {year} {2018})}\BibitemShut {NoStop}%
\bibitem [{\citenamefont {Giannozzi}\ \emph {et~al.}(2009)\citenamefont {Giannozzi}, \citenamefont {Baroni}, \citenamefont {Bonini}, \citenamefont {Calandra}, \citenamefont {Car}, \citenamefont {Cavazzoni}, \citenamefont {Ceresoli}, \citenamefont {Chiarotti}, \citenamefont {Cococcioni}, \citenamefont {Dabo} \emph {et~al.}}]{giannozziQUANTUMESPRESSOModular2009}%
  \BibitemOpen
  \bibfield  {author} {\bibinfo {author} {\bibfnamefont {P.}~\bibnamefont {Giannozzi}}, \bibinfo {author} {\bibfnamefont {S.}~\bibnamefont {Baroni}}, \bibinfo {author} {\bibfnamefont {N.}~\bibnamefont {Bonini}}, \bibinfo {author} {\bibfnamefont {M.}~\bibnamefont {Calandra}}, \bibinfo {author} {\bibfnamefont {R.}~\bibnamefont {Car}}, \bibinfo {author} {\bibfnamefont {C.}~\bibnamefont {Cavazzoni}}, \bibinfo {author} {\bibfnamefont {D.}~\bibnamefont {Ceresoli}}, \bibinfo {author} {\bibfnamefont {G.~L.}\ \bibnamefont {Chiarotti}}, \bibinfo {author} {\bibfnamefont {M.}~\bibnamefont {Cococcioni}}, \bibinfo {author} {\bibfnamefont {I.}~\bibnamefont {Dabo}}, \emph {et~al.},\ }\bibfield  {title} {\bibinfo {title} {{{QUANTUM ESPRESSO}}: A modular and open-source software project for quantum simulations of materials},\ }\href {https://doi.org/10.1088/0953-8984/21/39/395502} {\bibfield  {journal} {\bibinfo  {journal} {Journal of Physics: Condensed Matter}\ }\textbf {\bibinfo {volume} {21}},\ \bibinfo {pages} {395502}
  (\bibinfo {year} {2009})}\BibitemShut {NoStop}%
\bibitem [{\citenamefont {Giannozzi}\ \emph {et~al.}(2017)\citenamefont {Giannozzi}, \citenamefont {Andreussi}, \citenamefont {Brumme}, \citenamefont {Bunau}, \citenamefont {Buongiorno~Nardelli}, \citenamefont {Calandra}, \citenamefont {Car}, \citenamefont {Cavazzoni}, \citenamefont {Ceresoli}, \citenamefont {Cococcioni} \emph {et~al.}}]{giannozziAdvancedCapabilitiesMaterials2017}%
  \BibitemOpen
  \bibfield  {author} {\bibinfo {author} {\bibfnamefont {P.}~\bibnamefont {Giannozzi}}, \bibinfo {author} {\bibfnamefont {O.}~\bibnamefont {Andreussi}}, \bibinfo {author} {\bibfnamefont {T.}~\bibnamefont {Brumme}}, \bibinfo {author} {\bibfnamefont {O.}~\bibnamefont {Bunau}}, \bibinfo {author} {\bibfnamefont {M.}~\bibnamefont {Buongiorno~Nardelli}}, \bibinfo {author} {\bibfnamefont {M.}~\bibnamefont {Calandra}}, \bibinfo {author} {\bibfnamefont {R.}~\bibnamefont {Car}}, \bibinfo {author} {\bibfnamefont {C.}~\bibnamefont {Cavazzoni}}, \bibinfo {author} {\bibfnamefont {D.}~\bibnamefont {Ceresoli}}, \bibinfo {author} {\bibfnamefont {M.}~\bibnamefont {Cococcioni}}, \emph {et~al.},\ }\bibfield  {title} {\bibinfo {title} {Advanced capabilities for materials modelling with {{Quantum ESPRESSO}}},\ }\href {https://doi.org/10.1088/1361-648X/aa8f79} {\bibfield  {journal} {\bibinfo  {journal} {Journal of Physics: Condensed Matter}\ }\textbf {\bibinfo {volume} {29}},\ \bibinfo {pages} {465901} (\bibinfo {year}
  {2017})}\BibitemShut {NoStop}%
\bibitem [{\citenamefont {Dal~Corso}(2014)}]{dalcorsoPseudopotentialsPeriodicTable2014}%
  \BibitemOpen
  \bibfield  {author} {\bibinfo {author} {\bibfnamefont {A.}~\bibnamefont {Dal~Corso}},\ }\bibfield  {title} {\bibinfo {title} {Pseudopotentials periodic table: {{From H}} to {{Pu}}},\ }\href {https://doi.org/10.1016/j.commatsci.2014.07.043} {\bibfield  {journal} {\bibinfo  {journal} {Computational Materials Science}\ }\textbf {\bibinfo {volume} {95}},\ \bibinfo {pages} {337} (\bibinfo {year} {2014})}\BibitemShut {NoStop}%
\bibitem [{\citenamefont {Bellaiche}\ and\ \citenamefont {Vanderbilt}(2000)}]{bellaicheVirtualCrystalApproximation2000}%
  \BibitemOpen
  \bibfield  {author} {\bibinfo {author} {\bibfnamefont {L.}~\bibnamefont {Bellaiche}}\ and\ \bibinfo {author} {\bibfnamefont {D.}~\bibnamefont {Vanderbilt}},\ }\bibfield  {title} {\bibinfo {title} {Virtual crystal approximation revisited: {{Application}} to dielectric and piezoelectric properties of perovskites},\ }\href {https://doi.org/10.1103/PhysRevB.61.7877} {\bibfield  {journal} {\bibinfo  {journal} {Physical Review B}\ }\textbf {\bibinfo {volume} {61}},\ \bibinfo {pages} {7877} (\bibinfo {year} {2000})}\BibitemShut {NoStop}%
\bibitem [{\citenamefont {Marzari}\ \emph {et~al.}(2012)\citenamefont {Marzari}, \citenamefont {Mostofi}, \citenamefont {Yates}, \citenamefont {Souza},\ and\ \citenamefont {Vanderbilt}}]{marzariMaximallyLocalizedWannier2012a}%
  \BibitemOpen
  \bibfield  {author} {\bibinfo {author} {\bibfnamefont {N.}~\bibnamefont {Marzari}}, \bibinfo {author} {\bibfnamefont {A.~A.}\ \bibnamefont {Mostofi}}, \bibinfo {author} {\bibfnamefont {J.~R.}\ \bibnamefont {Yates}}, \bibinfo {author} {\bibfnamefont {I.}~\bibnamefont {Souza}},\ and\ \bibinfo {author} {\bibfnamefont {D.}~\bibnamefont {Vanderbilt}},\ }\bibfield  {title} {\bibinfo {title} {Maximally localized {{Wannier}} functions: {{Theory}} and applications},\ }\href {https://doi.org/10.1103/RevModPhys.84.1419} {\bibfield  {journal} {\bibinfo  {journal} {Reviews of Modern Physics}\ }\textbf {\bibinfo {volume} {84}},\ \bibinfo {pages} {1419} (\bibinfo {year} {2012})}\BibitemShut {NoStop}%
\bibitem [{\citenamefont {Tsirkin}(2021)}]{tsirkinHighPerformanceWannier2021}%
  \BibitemOpen
  \bibfield  {author} {\bibinfo {author} {\bibfnamefont {S.~S.}\ \bibnamefont {Tsirkin}},\ }\bibfield  {title} {\bibinfo {title} {High performance {{Wannier}} interpolation of {{Berry}} curvature and related quantities with {{WannierBerri}} code},\ }\href {https://doi.org/10.1038/s41524-021-00498-5} {\bibfield  {journal} {\bibinfo  {journal} {npj Computational Materials}\ }\textbf {\bibinfo {volume} {7}},\ \bibinfo {pages} {33} (\bibinfo {year} {2021})}\BibitemShut {NoStop}%
\bibitem [{\citenamefont {Gradhand}\ \emph {et~al.}(2012)\citenamefont {Gradhand}, \citenamefont {Fedorov}, \citenamefont {Pientka}, \citenamefont {Zahn}, \citenamefont {Mertig},\ and\ \citenamefont {Gy{\"o}rffy}}]{gradhandFirstprincipleCalculationsBerry2012a}%
  \BibitemOpen
  \bibfield  {author} {\bibinfo {author} {\bibfnamefont {M.}~\bibnamefont {Gradhand}}, \bibinfo {author} {\bibfnamefont {D.~V.}\ \bibnamefont {Fedorov}}, \bibinfo {author} {\bibfnamefont {F.}~\bibnamefont {Pientka}}, \bibinfo {author} {\bibfnamefont {P.}~\bibnamefont {Zahn}}, \bibinfo {author} {\bibfnamefont {I.}~\bibnamefont {Mertig}},\ and\ \bibinfo {author} {\bibfnamefont {B.~L.}\ \bibnamefont {Gy{\"o}rffy}},\ }\bibfield  {title} {\bibinfo {title} {First-principle calculations of the {{Berry}} curvature of {{Bloch}} states for charge and spin transport of electrons},\ }\href {https://doi.org/10.1088/0953-8984/24/21/213202} {\bibfield  {journal} {\bibinfo  {journal} {Journal of Physics: Condensed Matter}\ }\textbf {\bibinfo {volume} {24}},\ \bibinfo {pages} {213202} (\bibinfo {year} {2012})}\BibitemShut {NoStop}%
\bibitem [{\citenamefont {Ikeda}\ \emph {et~al.}(2020)\citenamefont {Ikeda}, \citenamefont {Tsunoda}, \citenamefont {Oogane}, \citenamefont {Oh}, \citenamefont {Morita},\ and\ \citenamefont {Ando}}]{ikedaFabricationEvaluationHighly2020}%
  \BibitemOpen
  \bibfield  {author} {\bibinfo {author} {\bibfnamefont {T.}~\bibnamefont {Ikeda}}, \bibinfo {author} {\bibfnamefont {M.}~\bibnamefont {Tsunoda}}, \bibinfo {author} {\bibfnamefont {M.}~\bibnamefont {Oogane}}, \bibinfo {author} {\bibfnamefont {S.}~\bibnamefont {Oh}}, \bibinfo {author} {\bibfnamefont {T.}~\bibnamefont {Morita}},\ and\ \bibinfo {author} {\bibfnamefont {Y.}~\bibnamefont {Ando}},\ }\bibfield  {title} {\bibinfo {title} {Fabrication and evaluation of highly c-plane oriented {{Mn3Sn}} thin films},\ }\href {https://doi.org/10.1063/1.5130459} {\bibfield  {journal} {\bibinfo  {journal} {AIP Advances}\ }\textbf {\bibinfo {volume} {10}},\ \bibinfo {pages} {015310} (\bibinfo {year} {2020})}\BibitemShut {NoStop}%
\bibitem [{\citenamefont {Xie}\ \emph {et~al.}(2022{\natexlab{b}})\citenamefont {Xie}, \citenamefont {Chen}, \citenamefont {Zhang}, \citenamefont {Mu}, \citenamefont {Zhang}, \citenamefont {Yan},\ and\ \citenamefont {Wu}}]{xieMagnetizationSwitchingPolycrystalline2022a}%
  \BibitemOpen
  \bibfield  {author} {\bibinfo {author} {\bibfnamefont {H.}~\bibnamefont {Xie}}, \bibinfo {author} {\bibfnamefont {X.}~\bibnamefont {Chen}}, \bibinfo {author} {\bibfnamefont {Q.}~\bibnamefont {Zhang}}, \bibinfo {author} {\bibfnamefont {Z.}~\bibnamefont {Mu}}, \bibinfo {author} {\bibfnamefont {X.}~\bibnamefont {Zhang}}, \bibinfo {author} {\bibfnamefont {B.}~\bibnamefont {Yan}},\ and\ \bibinfo {author} {\bibfnamefont {Y.}~\bibnamefont {Wu}},\ }\bibfield  {title} {\bibinfo {title} {Magnetization switching in polycrystalline {{Mn3Sn}} thin film induced by self-generated spin-polarized current},\ }\href {https://doi.org/10.1038/s41467-022-33345-2} {\bibfield  {journal} {\bibinfo  {journal} {Nature Communications}\ }\textbf {\bibinfo {volume} {13}},\ \bibinfo {pages} {5744} (\bibinfo {year} {2022}{\natexlab{b}})}\BibitemShut {NoStop}%
\end{thebibliography}%

\end{document}